\documentclass[aps,pre,reprint,superscriptaddress]{revtex4-2}
\usepackage{eurosym}
\usepackage{graphicx}
\usepackage{amsmath}
\usepackage{latexsym}
\usepackage{amsthm, amscd, amsfonts, amssymb}
\usepackage{multirow}
\usepackage{makecell}
\usepackage{dcolumn}
\usepackage{tabularx}
\usepackage{xcolor}
\usepackage{etoolbox}
\usepackage{threeparttable}

\begin{document}

\title{Beating stripe solitons arising from helicoidal spin-orbit coupling
in Bose-Einstein condensates}

\author{Cui-Cui Ding}
\affiliation{Research Group of Nonlinear Optical Science and Technology, Research Center of Nonlinear Science,\\ School of Mathematical and Physical Sciences, Wuhan Textile University, Wuhan 430200, China}
\author{Qin Zhou}
\email{qinzhou@whu.edu.cn}
\affiliation{Research Group of Nonlinear Optical Science and Technology, Research Center of Nonlinear Science,\\ School of Mathematical and Physical Sciences, Wuhan Textile University, Wuhan 430200, China}
\author{B. A. Malomed}
\affiliation{Department of Physical Electronics, School of Electrical Engineering,
	Faculty of Engineering, and the Center for Light-Matter University, Tel Aviv
	University, Tel Aviv, Israel}
\affiliation{Instituto de Alta Investigaci\'{o}n, Universidad de Tarapac\'{a},
Casilla 7D, Arica, Chile}

\date{\today }

\begin{abstract}
We demonstrate that the model of a spatially non-uniform two-component
Bose-Einstein condensate (BEC) featuring the helicoidal spin-orbit coupling
(SOC), gives rise to dark-bright soliton complexes characterized by
spatiotemporal periodic oscillations in each component. These solitons are
formed by the superposition of dark and bright ones, and exhibit a beating
state over time and a striped state across space, earning them the
designation of beating stripe solitons. Our analysis demonstrates that
helicoidal SOC significantly affects the formation and dynamical
properties of these solitons, also serving as the primary driver for spin
oscillations. Through the nonlinear superposition of the beating stripe
solitons, a range of intricate scenarios of the interaction between multiple
solitons emerges, including head-on collisions, bound states, and parallel
states.
\end{abstract}

\maketitle

\affiliation{Research Group of Nonlinear Optical Science and Technology,
Research Center of Nonlinear Science,\\ School of Mathematical and Physical
Sciences, Wuhan Textile University, Wuhan 430200, China}

\affiliation{Research Group of Nonlinear Optical Science and Technology,
Research Center of Nonlinear Science,\\ School of Mathematical and Physical
Sciences, Wuhan Textile University, Wuhan 430200, China}

\affiliation{Department of Physical Electronics, School of Electrical Engineering,
	Faculty of Engineering, and the Center for Light-Matter University, Tel Aviv
	University, Tel Aviv, Israel}
\affiliation{Instituto de Alta Investigaci\'{o}n, Universidad de Tarapac\'{a},
Casilla 7D, Arica, Chile}











\section{Introduction}


Vector solitons in multi-component systems have drawn much interest in
experimental and theoretical studies due to the great variety of their
shapes and properties~\cite%
{Qu2016,Farolfi2020,Chai2020,Kengne2021,Meng2022,Mao2022}. Owing to their
high degree of tunability, multi-component Bose-Einstein condensates (BECs)
offer an ideal platform for exploring novel types of vector solitons~\cite%
{Kevrekidis2008}. In particular, the Manakov system~\cite{Manakov1973}, as
the vectorial extension of the nonlinear Schr\"{o}dinger equation, models
diverse wave phenomena in fields such as BECs, nonlinear optics, and fluid
mechanics, including bright-bright~\cite{Zhang2009}, dark-bright \cite%
{Nistazakis2008,Majed2017,Yan2011}, and dark-dark solitons~\cite{Ling2015}.
In particular, dark-bright solitons have been studied in detail
theoretically\cite%
{Busch2001,Yin2011,Vijayajayanthi2008,Achilleos2011,Achilleos2012,Karamatskos2015,Katsimiga2018} and
created in the experiment \cite{Becker2008,Chen1996,Hamner2011,Yan2011}.

The nonlinear superposition of vector solitons typically results in complex
coherent or incoherent multi-soliton complexes~\cite%
{Akhmediev1998,Akhmediev1999,Akhmediev2000}, such as non-degenerate solitons~%
\cite{Stalin2019,Ramakrishnan2020,Qin2021} and beating solitons~\cite%
{Yan2012,Charalampidis2015,Park2000,Zhao2018,Li2023,Liu2024,Achilleos2014,Hamner2011}%
. By utilizing the internal SU(2) symmetry of the system, exact
vector-soliton states -- specifically, beating ones -- can be generated
through the superposition of dark-bright solitons~\cite%
{Li2023,Liu2024,Achilleos2014}. The density distribution of each component
of the beating soliton exhibits periodic breathing oscillations, while the
total density remains constant~\cite{Park2000,Liu2024}. Beating solitons
have been experimentally observed in two-component BECs~\cite{Hoefer2011}.
Numerical simulations have recently shown that the beating solitons exist
even in non-integrable systems~\cite{Charalampidis2016,Wang2021}. Several
new types of such solutions have also been discovered in the self-focusing
Manakov model~\cite{Gelash2023}.

When parameters of a continuous medium are subject to spatially periodic
modulation, the translational invariance is broken, and nonlinear localized
excitations usually cannot propagate freely~\cite{Morsch2006,Kartashov2011}.
Several methods have been proposed to support stable moving solitons in
periodic media. However, over sufficiently long propagation distances,
radiation losses become significant, leading to gradual destruction of the
solitons \cite{Sterke1988,Konotop2002,Sakaguchi2005,Kartashov2009}.
Nevertheless, when a spatially modulated system obeys special symmetries,
freely moving nonlinear localized waves, free of radiation losses, persist
over indefinitely long distances~\cite{Kartashov2017,Kartashov2019,Lin2024}, similar
to the exact solutions of the Ablowitz-Ladik lattice \cite{AL} and some
other discrete models \cite{Barash}. For example, the existence and
stability of freely moving solitons in a spatially inhomogeneous BEC with
helicoidal spin-orbit coupling (SOC) have been reported~\cite{Kartashov2017}.

In this paper, we investigate dark-bright soliton complexes featuring
spatiotemporal periodic oscillations in a non-uniform two-component BEC
under the action of helicoidal SOC. Establishing a gauge equivalence with
the Manakov system, we utilize the superposition of bright and dark solitons
of the latter system to construct soliton solutions of the former one that
exhibit periodic oscillations in both spatial and temporal directions. The
density distribution in each individual component of those solitons
simultaneously features beating and striped states, whereas the overall
density distribution of the system does not exhibit oscillations, therefore,
we term these solutions beating stripe solitons (BSSs). We conduct an
in-depth analysis of the structure and dynamical behavior of the BSSs,
identify existence parameter ranges for different types of solitons, and
investigate regularizing effects of the helicoidal SOC on these solitons.
Through the analysis of the properties of these solitons, including the
particle-density and spin-density distributions, we uncover spin oscillation
phenomena. Additionally, we examine interactions between different BSSs,
including head-on collisions and their bound and parallel states.

The subsequent presentation is arranged as follows. The model and its
general BSS solutions are presented in Sec. II. In Sec. III, we provide a
comprehensive analysis of the generation conditions and dynamical properties
of different types of BSSs, along with their three degenerate forms.
Specific analysis is presented for the regularizing effects of the
helicoidal SOC on the solitons and their relation with spin oscillations.
The interactions between BSSs, including the head-on collisions between
solitons with different velocities, as well as the bound and parallel states
of the solitons with equal velocities, are the subject of Sec. IV.
Conclusions are drawn in Sec. V.

\section{Model and fundamental beating stripe soliton (BSS) solutions}

We consider the one-dimensional spin-$1/2$ BEC with helicoidal SOC, which is
governed, in the mean-field approximation, by the two-component
Gross-Pitaevskii (GP) equation, written in the scaled form~\cite%
{Kartashov2017}
\begin{equation}
i\frac{\partial \mathbf{\Psi }}{\partial t}=\frac{1}{2}Q^{2}(x)\mathbf{\Psi }%
-s(\mathbf{\Psi }^{\dag }\mathbf{\Psi })\mathbf{\Psi }.
\label{helicoidal SOC}
\end{equation}%
Here the spinor wave function is $\mathbf{\Psi }=(\Psi _{1},\Psi _{2})^{T}$,
$s=\pm 1$ corresponds to the attractive and repulsive interatomic
interactions, respectively, and the helicoidally molded SOC is represented
by the generalized momentum operator,
\begin{equation}
Q(x)=-i\partial /\partial x+\alpha \boldsymbol{\sigma }\cdot \mathbf{n}(x)
\label{Qx}
\end{equation}%
\cite{Zhang2013,Jimenez2015,Luo2016}, with the experimentally~tunable SOC
strength $\alpha $, vector $\boldsymbol{\sigma }=(\sigma _{1},\sigma
_{2},\sigma _{3})$ of the Pauli matrices, and the spatial modulation defined
as
\begin{equation}
\mathbf{n}(x)=(\cos (2\kappa x),\sin (2\kappa x),0),
\label{spatial modulation}
\end{equation}%
where $\kappa >0$ and $\kappa <0$ signify the right- and left-handed
helicity, respectively~\cite{Rechtsman2013,Samsonov2004,Burt2004}. Equation (%
\ref{helicoidal SOC}) with $\kappa =0$ correspond to the uniform
Rashba-Dresselhaus SOC~\cite{Dalibard2011}, and to the canonical Manakov
system~\cite{Manakov1973} in the case of $\alpha =0$.

Equation~(\ref{helicoidal SOC}) for wave function $\mathbf{\Psi }\left(
x,t\right) $ can be transformed into the standard Manakov system for another
spinor wave function, $\mathbf{u}\left( x,t\right) $. by substitution
\begin{equation}
\mathbf{\Psi =Tu},  \label{T}
\end{equation}%
with matrix $\mathbf{T}$ defined\ by Eq.~(\ref{trans1}) in Appendix A, with
the determinant $\left\vert \mathbf{T}\right\vert =$ $-1$.\ We emphasize
that, unlike the use of \textit{SU}(2) transformations to construct beating
solitons, used in previous works~\cite%
{Yan2012,Charalampidis2015,Park2000,Zhao2018,Li2023,Liu2024,Achilleos2014,Hamner2011}%
, the transformation matrix here is a function of $x$, rather than a
constant. This non-uniform transformation, which is determined by the
helicoidal SOC, is also the cause of the formation of beating solitons with
a striped structure. Therefore, starting from the plane wave as the
zero-order seed solution, we can use the transformation, along with the
Darboux transform for the Manakov's system~\cite{Guo2012}, to construct
fundamental (first-order) BSS solutions of Eq.~(\ref{helicoidal SOC}) in the
following compact form:
\begin{equation}
\begin{aligned} \Psi_1=&e^{-i\kappa x}(\nu_+u_1+\nu_-u_2),\\
\Psi_2=&e^{i\kappa x}(\nu_-u_1-\nu_+u_2), \end{aligned}
\label{dark-bright soliton}
\end{equation}%
where
\begin{equation}
\begin{aligned} \nu _{+}=&\text{sgn}(\alpha )\sqrt{\left( k_{\text{m}}-
\kappa \right) /\left( 2k_{\text{m}}\right) } \\ \nu _{-}=&\sqrt{\left(
k_{\text{m}}+ \kappa \right) /\left( 2k_{\text{m} }\right) }. \end{aligned}
\label{nu}
\end{equation}%
and the dark and bright components are
\begin{equation}
\begin{aligned} u_j=\rho_ju_{j0}u^{db}_j,~~(j=1,2) \end{aligned}  \label{u12}
\end{equation}%
with $u_{j0}$ being the plane waves
\begin{equation}
\begin{aligned}
u_{10}=&a\exp{i[(k_1-k_{\text{m}})x+(sa^2-\frac{1}{2}k_1^2)t]},\\
u_{20}=&\exp{i[(k_2+k_{\text{m}})x+(sa^2-\frac{1}{2}k_2^2)t]}. \end{aligned}
\label{plane wave}
\end{equation}%
These solutions depend on the amplitude ($a$), wavenumbers ($k_{1,2}$), the
effective momentum, $k_{\text{m}}=\sqrt{\alpha ^{2}+\kappa ^{2}}$, and
\begin{equation}
\begin{aligned}
\rho_1=&\sqrt{\frac{(k_1-\mu_1^*)(k_1-\mu_2^*)}{(k_1-\mu_1)(k_1-\mu_2)}},\\
\rho_2=&\frac{l_3}{\sqrt{l_1l_2}}\sqrt{-4\mu_{1I}\mu_{2I}}. \end{aligned}
\label{rho12}
\end{equation}%
Additionally
\begin{equation}
\begin{aligned}
u^{db}_1=&\frac{\cosh(\Theta_1+i\epsilon)+\delta\cos(\Theta_2+i\varepsilon)+%
\gamma_2\exp(C_1)}
{\cosh(\Theta_1)+\delta\cos(\Theta_2)+\gamma_1\exp(C_1)},\\
u^{db}_2=&\frac{\cosh(\Theta_3)}
{\cosh(\Theta_1)+\delta\cos(\Theta_2)+\gamma_1\exp(C_1)}, \end{aligned}
\label{db}
\end{equation}%
where
\begin{equation}
\begin{aligned}
\Theta_1=&(\mu_{2I}-\mu_{1I})x+(\mu_{1R}\mu_{1I}-\mu_{2R}\mu_{2I})t+
\log{\sqrt{\frac{|l_2|^2\mu_{1I}}{|l_1|^2\mu_{2I}}}},\\
\Theta_2=&(\mu_{1R}-\mu_{2R})x+\frac{1}{2}(\mu_{1I}^2-\mu_{1R}^2-\mu_{2I}^2+%
\mu_{2R}^2)t\\
&+\arg{\left[\frac{l_1^*(\mu_1-\mu_2^*)}{l_2^*(\mu_1-\mu_1^*)}\right]},\\
\Theta_3=&-\frac{i}{4}(\mu_1^*-\mu_2^*)[2x-(\mu_1^*+\mu_2^*)t]
+\log{\sqrt{\frac{l_2^*}{l_1^*}}},\\
C_1=&-(\mu_{1I}+\mu_{2I})x+(\mu_{1I}\mu_{1R}+\mu_{2I}\mu_{2R})t.
\end{aligned}
\end{equation}%
Here, $\ast $ stands for the complex conjugate, subscripts $R$ and $I$
representing the real and imaginary parts of complex parameters. The
relation between eigenvalues $\mu _{1,2}$ and spectral parameter $\lambda $
is provided by the quadratic equation:
\begin{equation}
\begin{aligned} \mu^2+(2\lambda-k_1)\mu-sa^2-2k_1\lambda=0. \end{aligned}
\label{quadratic equation}
\end{equation}%
Other symbols used in Eq.~(\ref{db}) are
\begin{equation}
\begin{aligned} \epsilon=&\arg{\left(\frac{k_1-\mu_1}{k_1-\mu_2}\right)},~~
\varepsilon=\log{\left|\frac{k_1-\mu_2}{k_1-\mu_1}\right|},\\
\delta=&\frac{\sqrt{|\mu_1-\mu_1^*||\mu_2-\mu_2^*|}}{|\mu_1-\mu_2^*|},\\
\gamma_1=&\frac{s|l_3|^2\sqrt{|\mu_1-\mu_1^*||\mu_2-\mu_2^*|}}
{4|l_1l_2|(\lambda^*-\lambda)},~~\gamma_2=\gamma_1/\rho_1. \end{aligned}
\end{equation}

Solutions~(\ref{dark-bright soliton}) describe the spatiotemporal
periodically oscillating combination of dark-bright solitons, which exhibits
periodic beating in the temporal direction due to the superposition of dark
and bright solitons. In the spatial direction, striped states are generated
due to the helicoidal SOC, hence we refer to them as BSSs. These solitons
are determined by several factors, including the initial background height $%
a $, SOC strength $\alpha $, helicity pitch $\kappa $, spectral parameter $%
\lambda $, coefficients $(l_{1},l_{2},l_{3})$ of the vector eigenfunction,
and the nonlinearity coefficient $s$. Note that the initial height $a$ can
be fixed as $a=1$ by means of a scale transformation.

These solitons encompass two species: the fundamental one, when either $%
l_{1} $ or $l_{2}$ is zero, and the composite form, with $l_{1,2}\neq 0$. A
comprehensive analysis of these solutions is presented in the subsequent
section. To reproduce the fundamental form of BSSs from solutions~(\ref%
{dark-bright soliton}), one needs, first, to transform the solution into an
expression with a common denominator, and subsequently set $l_{1}=0$ or $l_{2}=0$.

\begin{table*}
\caption{\label{table1}Pproperties of the BSSs in both the attractive
and repulsive regimes.}
\begin{threeparttable}
\begin{ruledtabular}
\begin{tabular}{ccccccc}
Regime & Condition & Velocity $V_g$ & Width $W$ & Beating period in $t$ & Stripe period in $x$ & Total density \\ \hline
$s=-1$ & $\lambda_{1I}<\sqrt{2}a/4$ & $k_1$ & $1/|\lambda_{1I}+\sqrt{\lambda_{1I}^2-sa^2}|$ & $2\pi/|k_{t1}|$\tnote{*} & $\pi/|k_{\text{m}}|$
& Dark soliton  \\
 & $\lambda_{1I}>\sqrt{2}a/4$ &  &  &  &
& Bright soliton  \\\vspace{3mm}
 & $\lambda_{1I}=\sqrt{2}a/4$ &  &  &  &
& Plane wave  \\
$s=1$ & $sa^2-\lambda_{1I}^2\geq0$ & $k_1-\sqrt{sa^2-\lambda_{1I}^2}$ & $1/|\lambda_{1I}|$ & $2\pi/|k_{t2}|$\tnote{†} & $2\pi/|2k_{\text{m}}-\sqrt{sa^2-\lambda_{1I}^2}|$
& Bright soliton \\
 & $sa^2-\lambda_{1I}^2\leq0$ & $k_1$ & $1/|\lambda_{1I}+\sqrt{\lambda_{1I}^2-sa^2}|$ & $2\pi/|k_{t1}|$\tnote{*} & $\pi/|k_{\text{m}}|$
&  \\
\end{tabular}
\begin{tablenotes}
  \item[*] $k_{t1}=-\frac{sa^2}{2}+\lambda_{1I}(\lambda_{1I}
+\sqrt{\lambda_{1I}^2-sa^2}).$
  \item[†] $k_{t2}=-\frac{sa^2}{2}+\lambda_{1I}^2+k_1\sqrt{sa^2-\lambda_{1I}^2}.$
\end{tablenotes}
\end{ruledtabular}
\end{threeparttable}
\end{table*}

\section{Dynamical properties of beating stripe solitons (BSSs)}

\subsection{The general structure of BSSs}

In this section, we first consider a simple case of solutions~(\ref%
{dark-bright soliton}) with $l_{1}$ or $l_{2}$ being zero. To simplify the
expressions for the solutions, we set
\begin{equation}
\begin{aligned} l_1=1,~~l_2=0,~~l_3=\sqrt{\frac{-2\lambda_{1I}}{s\mu_{1I}}},
\end{aligned}  \label{l123}
\end{equation}%
and reduce solution~(\ref{dark-bright soliton}) to
\begin{equation}
\begin{aligned} \Psi_1=&e^{-i\kappa x}(\nu_+\Psi_{DS}+\nu_-\Psi_{BS}),\\
\Psi_2=&e^{i\kappa x}(\nu_-\Psi_{DS}-\nu_+\Psi_{BS}), \end{aligned}
\label{Stripe dark-bright}
\end{equation}%
where the dark and bright solitons, $\Psi _{DS}$ and $\Psi _{BS}$, are
\begin{equation}
\begin{aligned} \Psi_{DS}=&a e^{i\theta^\prime_1}\left[1+\frac{i\mu_{1I}}
{k_1-\mu_1}(1+\tanh{\xi})\right],\\
\Psi_{BS}=&il_3\mu_{1I}e^{i\theta^\prime_2}~\text{sech}{\xi}, \end{aligned}
\label{d-b}
\end{equation}%
with
\begin{equation}
\begin{aligned}
\xi=&\mu_{1I}(x-\mu_{1R}t),~~\theta^\prime_1=-k_{\text{m}}x+\theta_1,\\
\theta^\prime_2=&(k_{\text{m}}-k_2+\mu_{1R})x+
\frac{1}{2}(k_2^2+\mu_{1I}^2-\mu_{1R}^2)t+\theta_2. \end{aligned}
\end{equation}

Both components $\Psi _{1}$ and $\Psi _{2}$ described by solutions~(\ref%
{Stripe dark-bright}) are formed by the superposition of the dark soliton $%
\Psi _{DS}$ and the bright one $\Psi _{BS}$, which triggers the emergence of
beating solitons on a nonvanishing background. Due to the SOC effect, these
solitons simultaneously exhibit the striped state, thus producing the stripe
solitons with the beating pattern. Hereafter, we refer to them as BSSs.

\begin{figure*}
\includegraphics[scale=0.35]{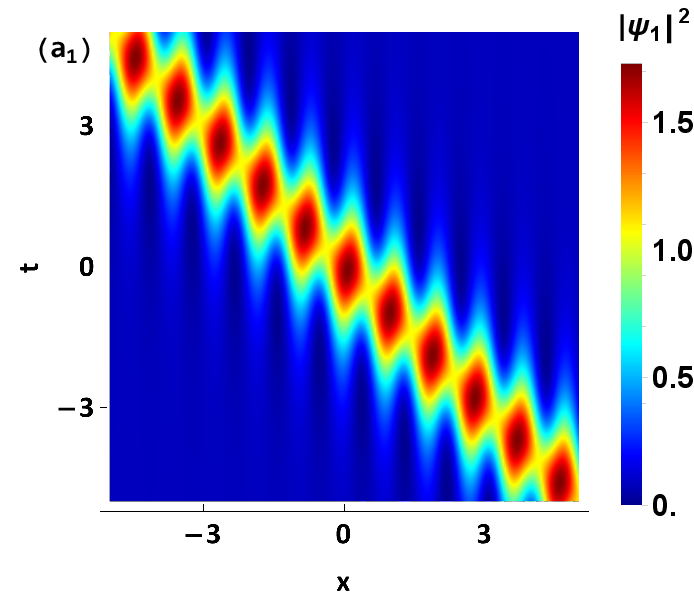}\hspace{3mm}
\includegraphics[scale=0.35]{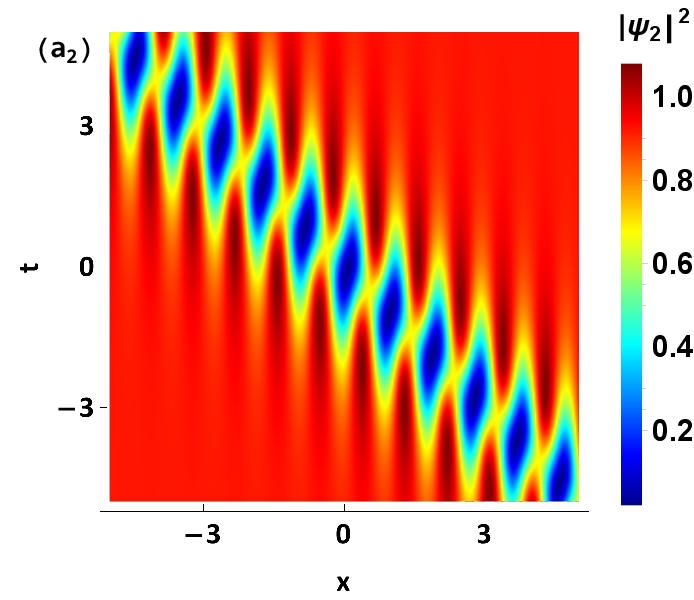} \hspace{3mm}
\includegraphics[scale=0.35]{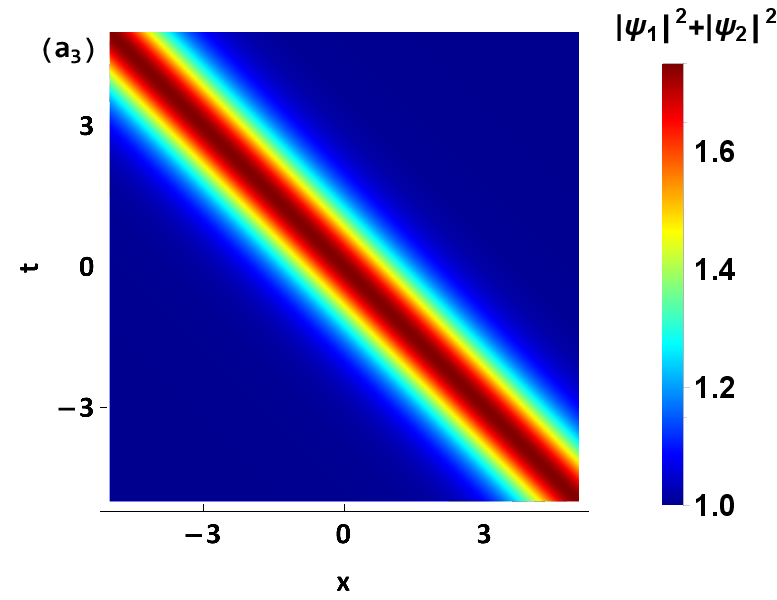} \hspace{3mm}\newline
\includegraphics[scale=0.35]{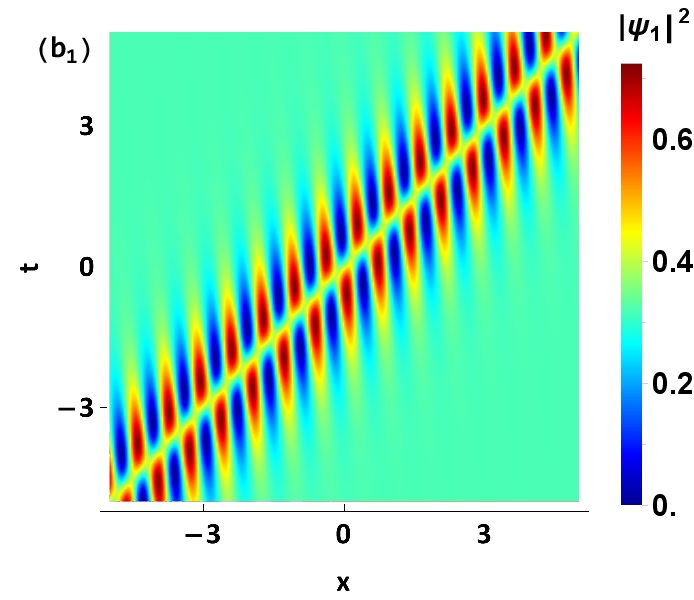}\hspace{3mm}
\includegraphics[scale=0.35]{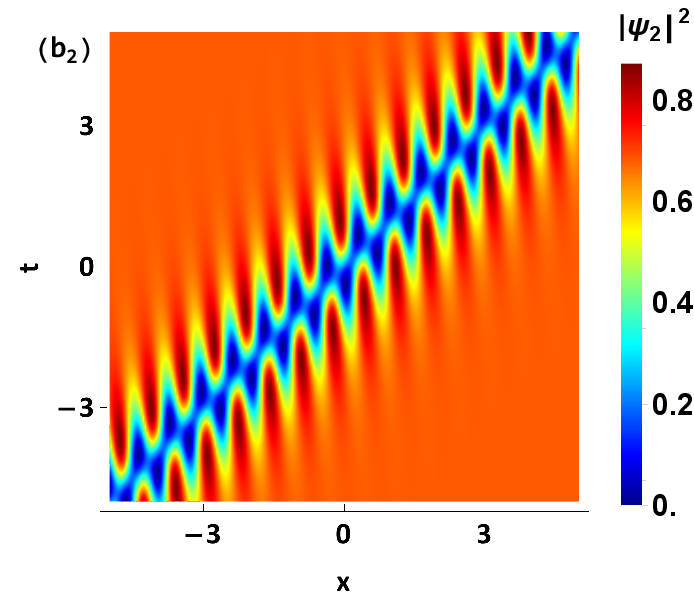} \hspace{3mm}
\includegraphics[scale=0.35]{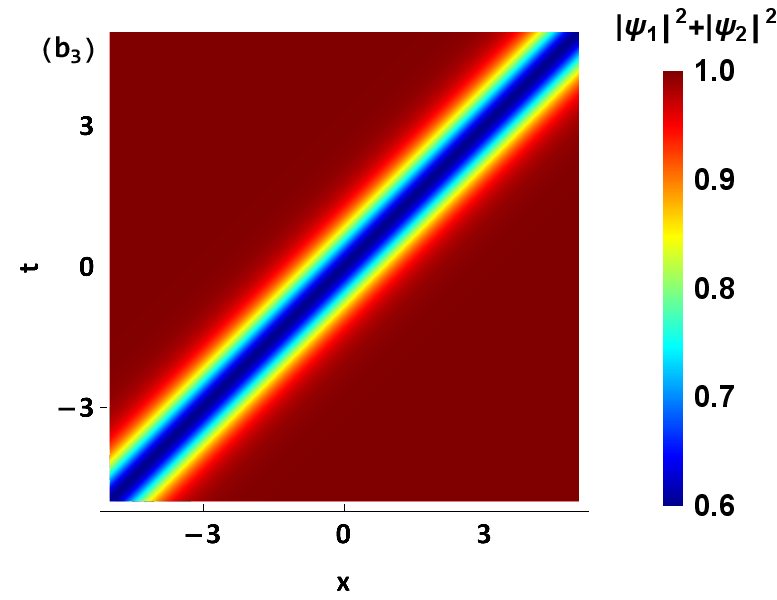} \hspace{3mm}\newline
\includegraphics[scale=0.35]{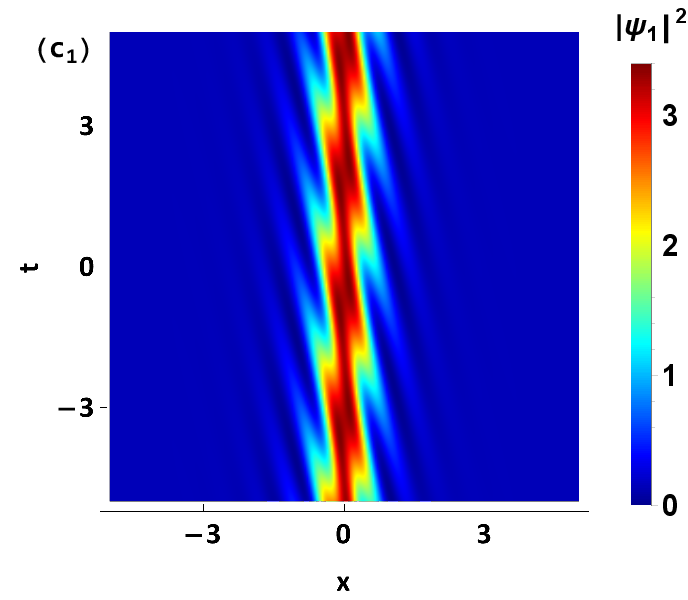}\hspace{3mm}
\includegraphics[scale=0.35]{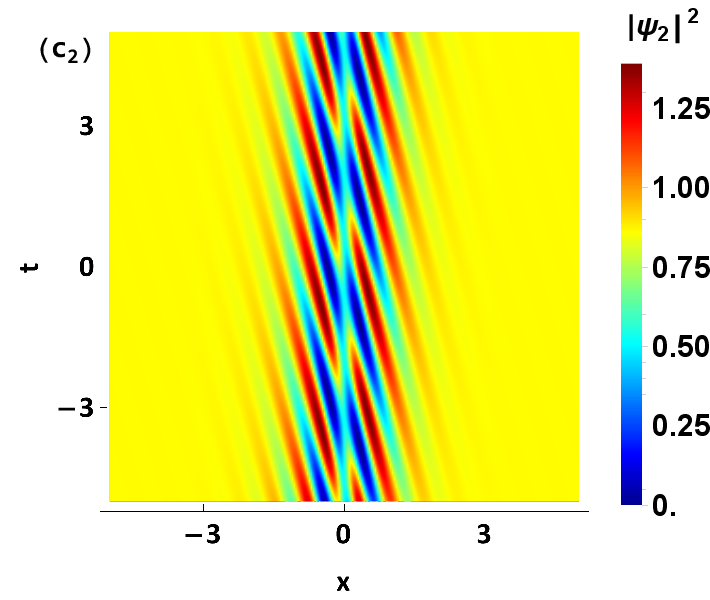} \hspace{3mm}
\includegraphics[scale=0.35]{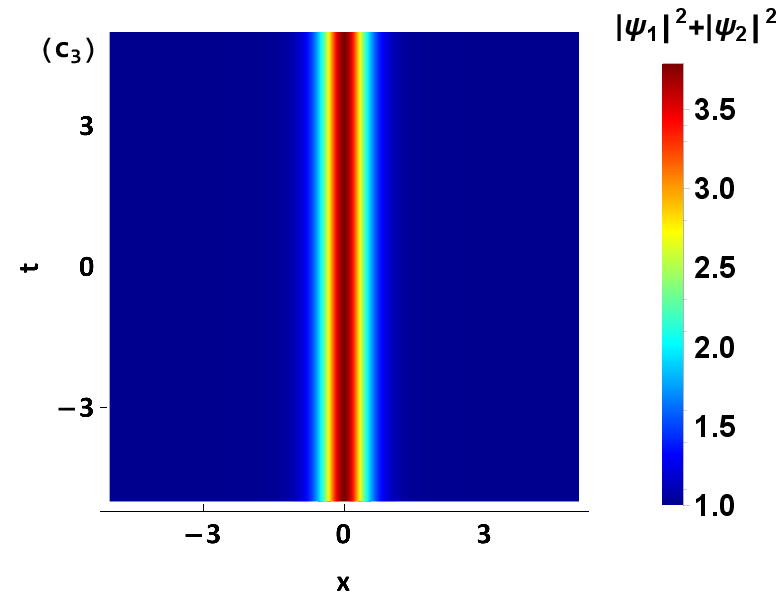} \hspace{3mm}\newline
\caption{\label{fig1}Beating stripe solitons with
different dynamic characteristics in both attraction and repulsion regimes in components $|\Psi
_{1}|^{2}$, $|\Psi _{2}|^{2}$ and the total density $|\Psi |^{2}=|\Psi
_{1}|^{2}+|\Psi _{2}|^{2}$, as produced by solutions (\protect\ref{Stripe
dark-bright}). $(a)$ A moving soliton in the attraction regime~$(s=1)$, with
$k_{1}=-\frac{1}{2}$, $\protect\lambda _{1}=\frac{1}{4}+\frac{\protect\sqrt{3%
}}{2}i$, $\protect\alpha =2$ and $\protect\kappa =\protect\pi $. $(b)$ A
moving soliton in the repulsion regime~$(s=-1)$, with $k_{1}=1$, $\protect%
\lambda _{1}=-\frac{1}{2}+\frac{\protect\sqrt{2}}{6}i$, $\protect\alpha =4$
and $\protect\kappa =\frac{\protect\pi }{2}$. $(c)$ A static soliton in the
repulsion regime~$(s=-1)$ with $k_{1}=0$, $\protect\lambda _{1}=\frac{%
\protect\sqrt{3}}{2}i$, $\protect\alpha =3$, and $\protect\kappa =\protect%
\pi $. The amplitude is $a=1$.}
\end{figure*}

To better reveal the dynamic properties and generation mechanism of these
BSSs, we first focus on their densities, which can be expressed
as
\begin{equation}
\begin{aligned}
|\Psi_1|^2=&a^2\nu_+^2+B_1~\text{sech}^2\xi+a\nu_+\nu_-\mu_{1I}\Psi_p~%
\text{sech}\xi,\\
|\Psi_2|^2=&a^2\nu_-^2+B_2~\text{sech}^2\xi-a\nu_+\nu_-\mu_{1I}\Psi_p~%
\text{sech}\xi, \end{aligned}  \label{modulus beating}
\end{equation}%
where
\begin{equation}
\begin{aligned}
&B_{1,2}=\mu_{1I}^2\left[\nu_{\mp}^2|l_3|^2-\frac{a\nu_{\pm}^2}{|k_1-%
\mu_1|^2}\right],\\ &\Psi_{p_1}=2|l_3|\left[\frac{\mu_{1I}
\cos(p_1+\varsigma_2)}{|k_1-\mu_1|}(1+\tanh\xi)-\sin(p_1+\varsigma_1)%
\right],\\ &\varsigma_1=\arg(l_3),~~\varsigma_2=\arg[l_3(k_1-\mu_1)],
\end{aligned}
\end{equation}%
and
\begin{equation}
\begin{aligned} p_1=k_x x+k_t t \end{aligned}  \label{periodic beating}
\end{equation}%
with
\begin{equation}
\begin{aligned} k_x=&2k_{\text{m}}-k_1+\mu_{1R},\\
k_t=&\frac{1}{2}(k_1^2+\mu_{1I}^2-\mu_{1R}^2), \end{aligned}  \label{kxkt}
\end{equation}%
which characterize the stripe structure and beating states of the solitons.
Due to the energy exchange between the two components, the solitons undergo
periodic oscillations in the $x$ and/or $t$ directions, which is represented
by the periodic expression $\Psi _{p_1}$.

Solving Eq.~(\ref{quadratic equation}), we obtained the eigenvalues
\begin{equation}
\begin{aligned}
\mu_{1,2}=\frac{k_1}{2}-\lambda_1\mp\frac{1}{2}\sqrt{4sa^2+(k_1+2%
\lambda_1)^2}. \end{aligned}  \label{mu12}
\end{equation}%
To simplify the analysis, we set $\lambda _{1R}=-\frac{k_{1}}{2}$ and $k_2=0$ by default hereafter.
The BSSs produced by solutions~(\ref{Stripe dark-bright})
move with the group velocity
\begin{equation}
\begin{aligned} V_g=\mu_{1R}, \end{aligned}  \label{group velocity}
\end{equation}%
which has the specific form $\mu _{1R}=k_{1}-\sqrt{sa^{2}-\lambda _{1I}^{2}}$
for $sa^{2}-\lambda _{1I}^{2}>0$, while $\mu _{1R}=k_{1}$ for $%
sa^{2}-\lambda _{1I}^{2}<0$. Based on the general definition of the soliton width and in conjunction with the definition
of the width for beating solitons presented in Ref.~\cite{Liu2024}, here we define the width of the BSSs as
\begin{equation}
\begin{aligned} W=1/|\mu_{1I}|, \end{aligned}  \label{Width}
\end{equation}
which can be written explicitly as $W=1/|\lambda _{1I}|$ for $sa^{2}-\lambda
_{1I}^{2}>0$ and $W=1/|\lambda _{1I}+\sqrt{\lambda _{1I}^{2}-sa^{2}}|$ for $%
sa^{2}-\lambda _{1I}^{2}<0$. We stress that $sa^{2}-\lambda _{1I}^{2}$ is
always negative when $s=-1$.

Unlike the periodic oscillations exhibited by the superposition of the dark
and bright solitons in a particular component, the total atomic density of
the two components does not oscillate. Indeed, using the single-component
density in Eq.~(\ref{modulus beating}), the total density $|\Psi |^{2}=|\Psi
_{1}|^{2}+|\Psi _{2}|^{2}$ can be found as
\begin{equation}
\begin{aligned}
|\Psi|^2=a^2+\mu_{1I}^2\left(|l_3|^2-\frac{a^2}{|k_1-\mu_1|^2}\right)%
\text{sech}^2\xi. \end{aligned}  \label{total modulus}
\end{equation}%
The identity relation $\nu _{+}^{2}+\nu _{-}^{2}\equiv 1$ is used to obtain
this total density. Equation~(\ref{total modulus}) shows that the total
density is shaped as a bell-shaped soliton with a nonzero background $a^{2}$%
, which possesses the same group velocity as the single components. The
superposition of the bright and dark solitons results in the two components
oscillating in opposite phases. Any protrusion (or depression) in component $%
\Psi _{1}$ mirrors as a depression (or protrusion) in $\Psi _{2}$,
maintaining a constant total density and thus giving rise to the BSS.

Dynamical properties of those BSSs in both attractive and repulsive regimes,
including their velocity, width, period, and total density, are given in
Table~\ref{table1}. From the table, one can obtain the dynamical
characteristics of the solitons under different parameter conditions.
Specifically, the stripe period of the solitons in the $x$ direction is
directly controlled by the SOC effect, whereas the beating period in the $t$
direction is only related to spectral parameters. Moreover, in the case of
repulsion, the distribution of the total density in the system can exhibit
dark/bright soliton forms or even a plane wave, while it always reveals the
bright soliton in the case of attraction.

\begin{figure*}
\includegraphics[scale=0.35]{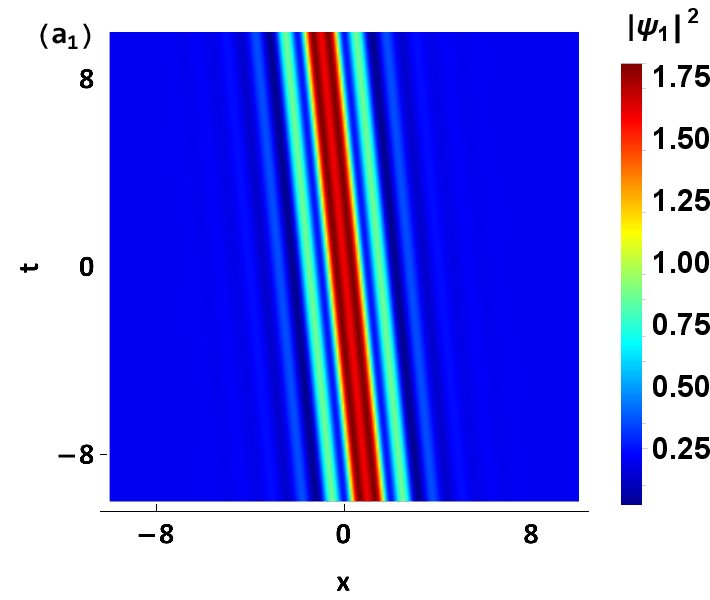}\hspace{3mm}
\includegraphics[scale=0.35]{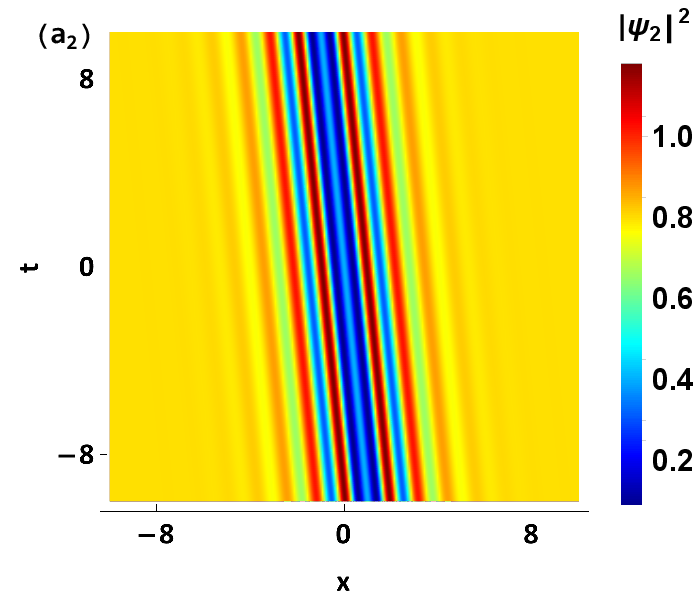} \hspace{3mm}
\includegraphics[scale=0.35]{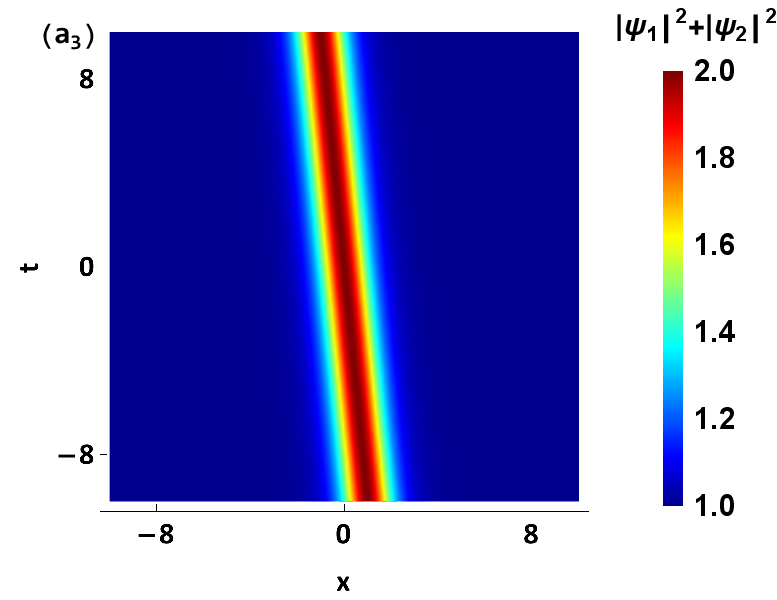} \hspace{3mm}\newline
\includegraphics[scale=0.35]{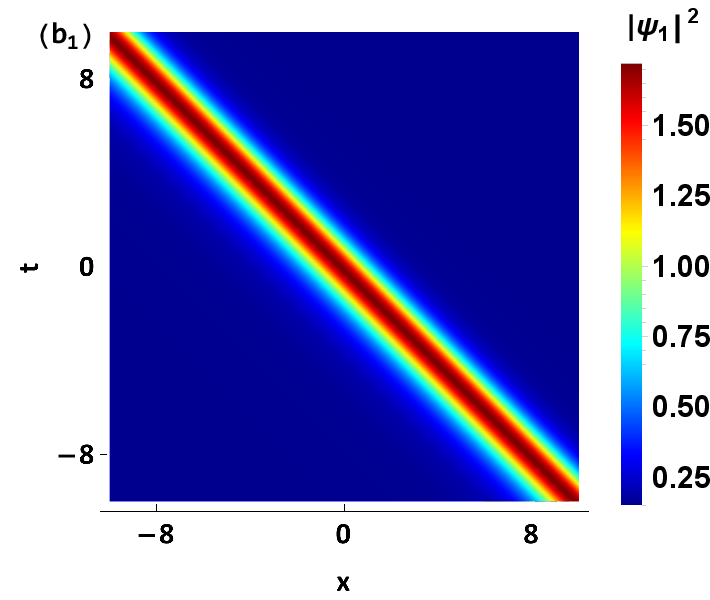}\hspace{3mm}
\includegraphics[scale=0.35]{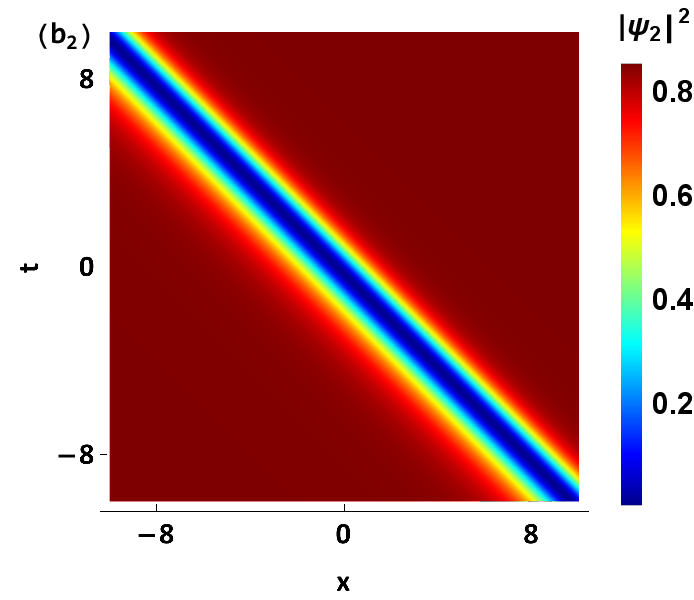} \hspace{3mm}
\includegraphics[scale=0.35]{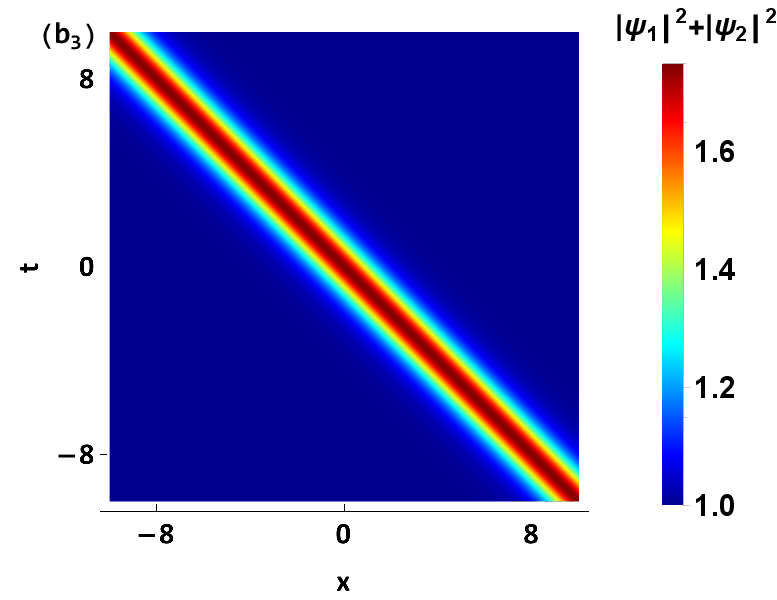} \hspace{3mm}\newline
\includegraphics[scale=0.35]{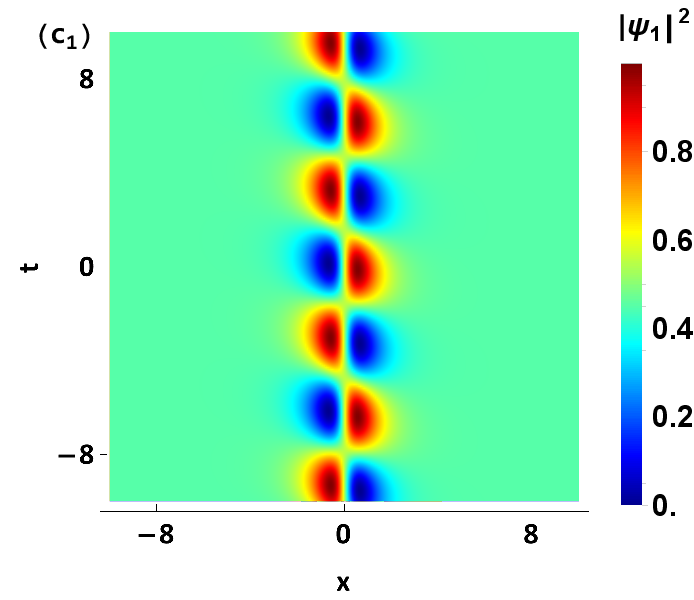}\hspace{3mm}
\includegraphics[scale=0.35]{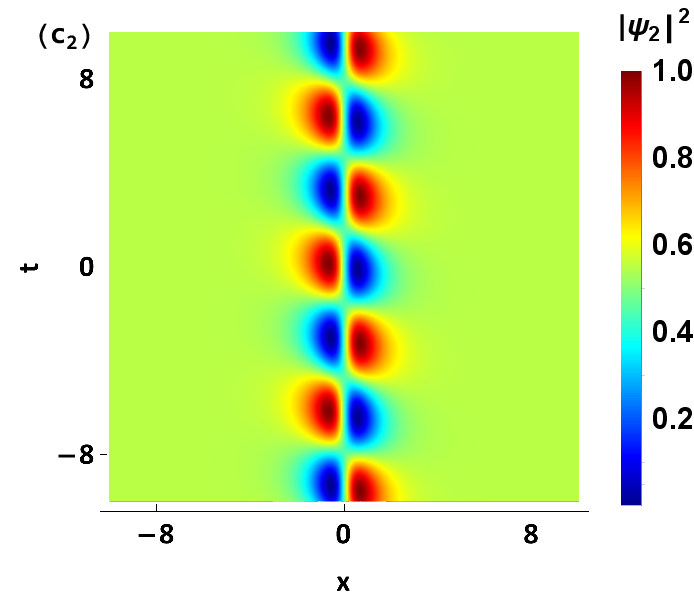} \hspace{3mm}
\includegraphics[scale=0.35]{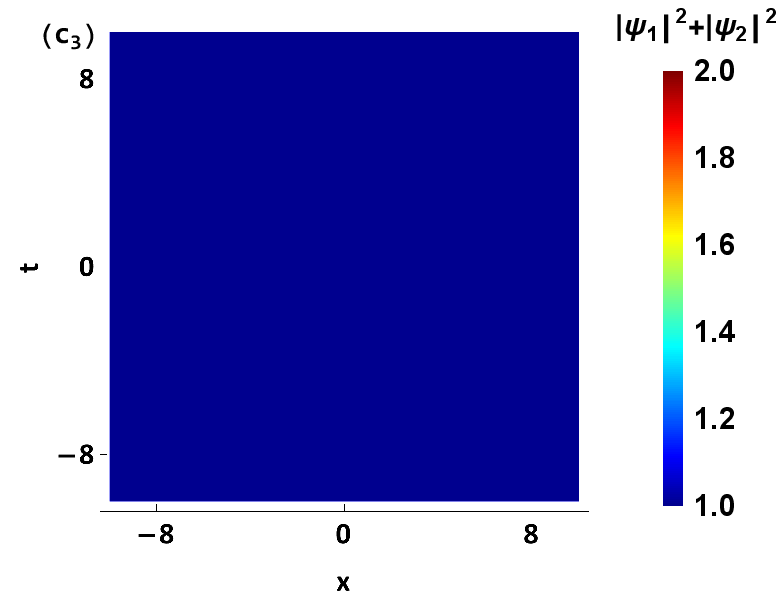} \hspace{3mm}\newline
\vspace{-3mm}
\caption{\label{fig2}Three degenerate beating solitons obtained from solutions (\protect
\ref{Stripe dark-bright}). $(a)$ A bright-dark pair with vibrations in the
attraction regime~$(s=1)$, with $k_{1}=-0.1$, $\protect\lambda _{1}=0.05-i$,
$\protect\alpha =2$, and $\protect\kappa =\frac{3}{2}$. $(b)$ A bright-dark
pair without vibrations in the attraction regime $(s=1)$, with $k_{1}=-\frac{%
1}{2}$, $\protect\lambda _{1}=\frac{1}{4}+\frac{\protect\sqrt{3}}{2}i$, and $%
\protect\alpha =\protect\kappa =\frac{\protect\sqrt{2}}{8}$. $(c)$
Complementary beating solitons with a constant total amplitude in the
repulson regime $(s=-1)$, with $k_{1}=0$, $\protect\lambda _{1}=\frac{%
\protect\sqrt{2}}{4}i$, $\protect\alpha =0.1$, and $\protect\kappa =0.01$.}
\end{figure*}

BSSs given by solutions~(\ref{Stripe dark-bright}) with
different dynamic characteristics in both attraction and repulsion regimes
are shown in Fig.~\ref{fig1}. Analysis of Eq.~(\ref{total modulus}) reveals
that, in the attraction regime $(s=1)$, the total amplitude of the two
components exceeds the background amplitude $a$ along the characteristic
line $\xi =0$, behaving as a bright soliton, as shown in Fig.~\ref{fig1}$(a)$%
. However, in the repulsion regime $(s=-1)$, for $\lambda _{1I}<\frac{\sqrt{2%
}}{4}a$, the total amplitude is less than the background amplitude,
exhibiting a dark soliton, as seen in Fig.~\ref{fig1}$(b)$, while for $%
\lambda _{1I}>\frac{\sqrt{2}}{4}a$ the total amplitude exceeds the
background value, looking as a bright soliton, as shown in Fig.~\ref{fig1}$%
(c)$.

Properly adjusting the spectral parameters, we can display BSSs that exhibit
diverse propagation characteristics: the backward propagation ($V_{g}<0$),
forward propagation ($V_{g}>0$), and even the stationary behavior ($V_{g}=0$%
), as shown in Figs.~\ref{fig1}$(a)$, $(b)$, and $(c)$, respectively.
Furthermore, depending on the type and count of extrema within a single
periodicity cycle of the BSS, solitons exhibiting distinct configurations
emerge, including bright-type solitons (featuring one maximum and two
minima), dark-type solitons (featuring one minimum and two maxima), and
four-petal solitons, with two maxima and two minima. An extensive variety of
combined soliton structures can be produced. As illustrations, we showcase a
bright-dark BSS (Fig.~\ref{fig1}$(a)$), a four-petal-four-petal BSS (Fig.~%
\ref{fig1}$(b)$), and a bright-four-petal BSS (Fig.~\ref{fig1}$(c)$).

\subsection{Control of beating stripe solitons (BSSs) by the helicoidal SOC}

SOC strength $\alpha $ and helicity pitch $\kappa $ do not alter the
propagation speed of the BSSs, but rather significantly affect their
structure. Making use of Eq.~(\ref{modulus beating}), one can develop a
comprehensive analysis of how the helicoidal SOC modulates the behavior of
the BSSs, which feature three types of the degeneration, as illustrated by
Fig.~\ref{fig2}. Two of the degenerate types manifest themselves as
bright-dark soliton pairs, with one exhibiting oscillations in the $x$
and/or $t$ directions, while the other remains oscillation-free in those
directions, as shown in Figs.~\ref{fig2}$(a)$ and $(b)$. The other type is
built as a pair of the complementary beating solitons, with the amplitude of
the two components equals the background amplitude $a$, as shown in Fig.~\ref%
{fig2}$(c)$. The formation of the oscillatory bright-dark soliton pairs
displayed in Fig.~\ref{fig2}$(a)$ occurs under the degenerate condition, in
which the soliton's characteristic line $\xi =0$ is parallel to the
oscillation direction $p_{1}=0$ -- specifically, at $k_{t}/k_{x}=-\mu _{1R}$%
, which is represented by parameters $k_{\text{m}}$ (including $\alpha $ and
$\kappa $), $\lambda _{1}$ and $k_{1}$. Conversely, the non-oscillatory
bright-dark soliton pairs displayed in Fig.~\ref{fig2}$(b)$ may solely
emerge when both $k_{x}$ and $k_{t}$ in periodic term~(\ref{periodic beating}%
) simultaneously vanish, in the specific case of $k_{\text{m}}\equiv \sqrt{%
\alpha ^{2}+\kappa ^{2}}=\sqrt{sa^{2}-\lambda _{1I}^{2}}/2$ and $k_{1}=\frac{%
sa^{2}-2\lambda _{1I}^{2}}{2\sqrt{sa^{2}-\lambda _{1I}^{2}}}$. Analysis
reveals that this degenerate case occurs solely in the attraction regime ($%
s=1$). To produce the two complementar BSSs shown in Fig.~\ref{fig2}$(c)$,
both $\lambda _{1I}=\frac{\sqrt{2}}{4}a$ and repulsion regime ($s=-1$) must
take place. Such a degenerate case is not possible in the attraction regime.
Specific parameter conditions for producing these three kinds of the
degenerate BSSs are given in Table~\ref{table2}.

\begin{table}[h]
\caption{Parameter conditions for producing the three kinds of degenerate
beating stripe solitons.}
\label{table2}%
\begin{ruledtabular}
\begin{tabular}{cc}
Three degenerate cases &  Conditions\\ \hline
\makecell[c]{Bright-dark soliton pair \\with vibration} & $k_t/k_x=-\mu_{1R}~(s=\pm1)$\\ \vspace{3mm}
\makecell[c]{Bright-dark soliton pair \\without vibration} & \makecell[c]{$k_{\text{m}}=\sqrt{sa^2-\lambda_{1I}^2}/2$,\\ $k_1=\frac{sa^2-2\lambda_{1I}^2}{2\sqrt{sa^2-\lambda_{1I}^2}}~~(s=1)$}\\ \vspace{3mm}
Complementary soliton & $\lambda_{1I}=\frac{\sqrt{2}}{4}a~~(s=-1)$\\
\end{tabular}
\end{ruledtabular}
\end{table}

\begin{figure}[tbp]
\includegraphics[scale=0.3]{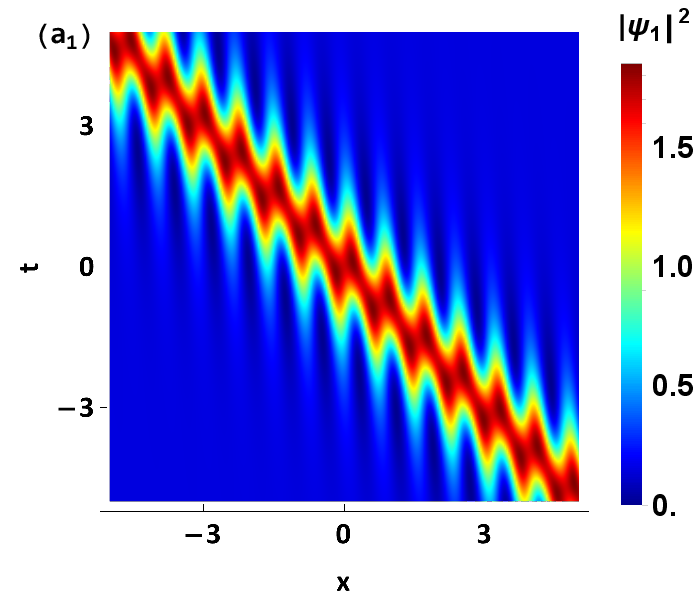}\hspace{3mm} %
\includegraphics[scale=0.3]{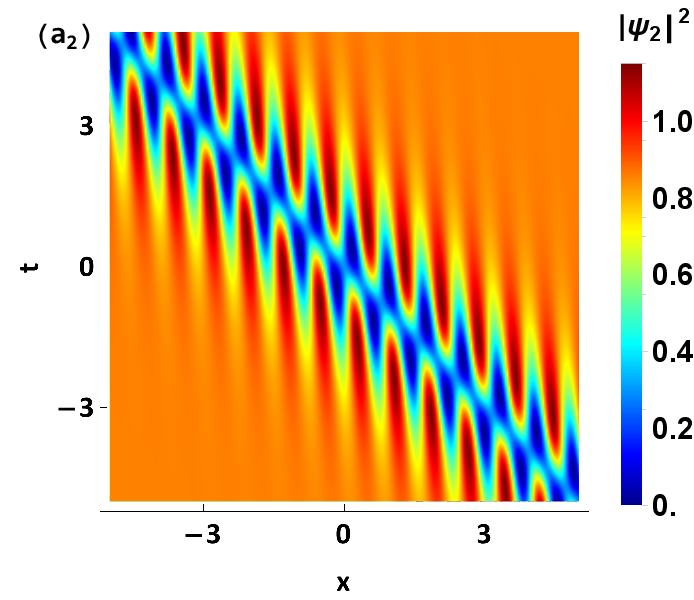}\newline
\includegraphics[scale=0.4]{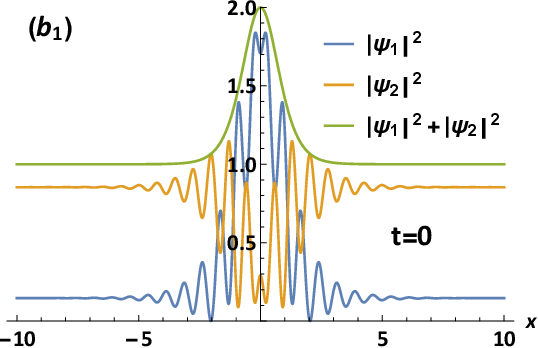}\hspace{3mm} %
\includegraphics[scale=0.4]{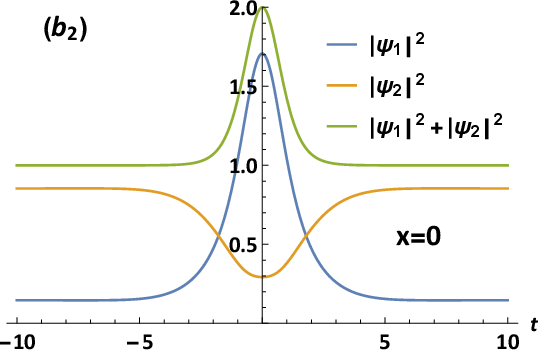}\newline
\vspace{3mm} \includegraphics[scale=0.65]{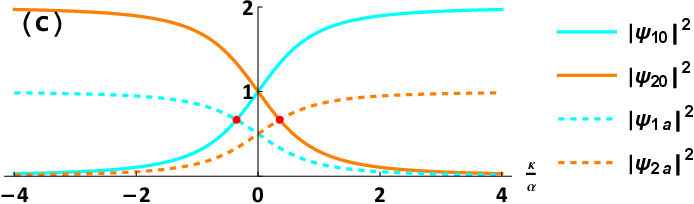}\newline
\vspace{-3mm}
\caption{$(a)$ and $(b)$ The density distribution of beating solitons with
symmetric stripes and their symmetric evolution along the directions of $t=0$
and $x=0,$ with $\protect\alpha =\protect\kappa =3$. $(c)$ The amplitudes of
the central stripes $|\Psi _{j0}|^{2}$ and background heights $|\Psi
_{ja}|^{2}$ of the two components vs.ratio $\frac{\protect\kappa }{\protect%
\alpha }$. The red dot corresponds to the transition between the bright and
dark solitons. The other parameters are $s=1$, $a=1$, $k_{1}=-1$, and $%
\protect\lambda _{1}=\frac{1}{2}-i$.}
\label{fig3}
\end{figure}

The SOC effect modulates the spatial oscillation frequency of the beating
solitons, resulting in the formation of stripe states, as is evident from
Fig.~\ref{fig3}. The modulation period in the spatial $x$ direction is $2\pi
/k_{x}$. Typically, the stripes of beating solitons exhibit asymmetry with
respect to the $x$- and $t$-axes. However, under specific conditions, a
symmetric configuration can emerge. Our analysis reveals that the beating
solitons with symmetric stripes are exclusively observed in attraction
regime $(s=1)$ under condition
\begin{equation}
\begin{aligned} \lambda_{1I}\leq-|a|. \end{aligned}
\label{symmetric stripe}
\end{equation}%
Figures~\ref{fig3}$(a)$ and $(b)$ showcase a remarkable example of such a
bright-dark beating soliton with symmetric stripes. In this system, one
component manifests itself as a bright soliton, while the other assumes is a
dark soliton. The character of each soliton component is determined by the
relative magnitude of SOC intensity $\alpha $ and rotation frequency $\kappa
$ (actually, by their ratio $\varpi =\frac{\kappa }{\alpha }$, as shown in
Fig.~\ref{fig3}$(c)$. Under a certain symmetry condition, \textit{viz}., $%
s=1 $, $a=1$ and $\lambda _{1I}=-1$, the densities of the two components at $%
(x,t)=(0,0)$ are calculated as
\begin{equation}
\begin{aligned} |\Psi_{10}|^2=&1+\frac{\varpi}{\sqrt{1+\varpi^2}},\\
|\Psi_{20}|^2=&1-\frac{\varpi}{\sqrt{1+\varpi^2}}, \end{aligned}
\label{varpi}
\end{equation}%
with background heights $|\Psi _{1a}|^{2}=\left[ 1+(\varpi +\sqrt{1+\varpi
^{2}})^{2}\right] ^{-1}$ and $|\Psi _{2a}|^{2}=1-|\Psi _{1a}|^{2}$, Their
evolution of with the change of ratio $\frac{\kappa }{\alpha }$ is
represented in Fig.~\ref{fig3}$(c)$. By comparing these values, one can
unambiguously classify the soliton as either bright or dark one. Remarkably,
the critical threshold for the transition between the bright and dark
solitons is identified as $\varpi =\pm \frac{\sqrt{2}}{4}$ (red points in
Fig.~\ref{fig3}$(c)$). When $\varpi >\frac{\sqrt{2}}{4}$, we observe a
bright-dark BSS. Conversely, when $\varpi <-\frac{\sqrt{2}}{4}$, a
dark-bright BSS emerges. The intermediate range represents a transition
state, where the solitons exhibit a blend of bright and dark characteristics.

\begin{figure*}
\includegraphics[scale=0.45]{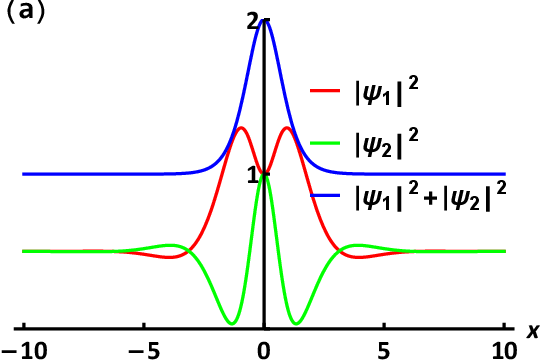}\hspace{3mm}
\includegraphics[scale=0.45]{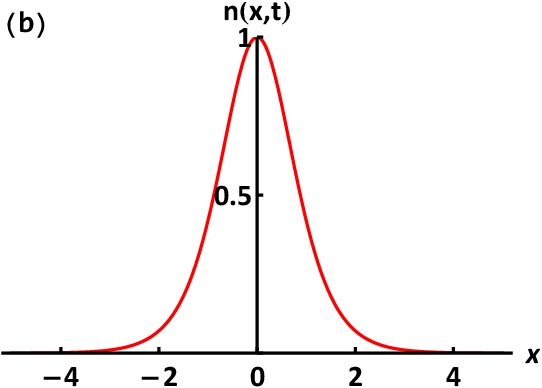} \hspace{3mm}
\includegraphics[scale=0.45]{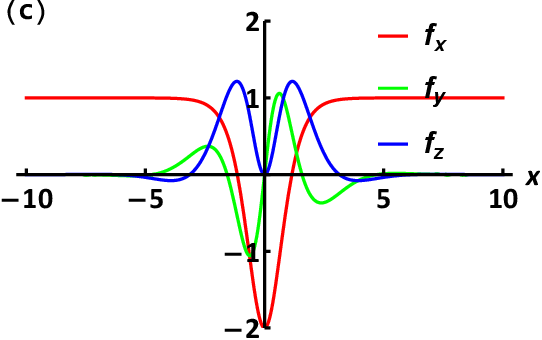} \hspace{3mm}
\vspace{3mm}
\caption{\label{fig30}Physical properties of the wave functions for the beating stripe
solitons at instant $t=0$. $(a)$ The local density; $(b)$ The particle
density; $(c)$ The spin density. The parameters are $s=1$, $a=1$, $k_{1}=0$,
$\protect\lambda _{1}=-i$, $\protect\alpha =0.5$, and $\protect\kappa =0$.}
\end{figure*}

To complete the discussion on general BSSs, we finally consider two special cases: (i) $\kappa=0$ but $\alpha\neq0$  but $\alpha\neq0$
(corresponding to the Rashba-Dresselhaus SOC), and (ii) $\alpha=0$ (corresponding to the conventional BEC). For the first case, from the density distribution of the two components given by Eq.~(\ref{modulus beating})
in conjunction with Eqs.~(\ref{periodic beating}) and (\ref{kxkt}), it can be seen that the fact whether $\kappa$  is zero or not does not fundamentally affect the structure of the BSSs; however, its sign (corresponding to the right- and left-handed helicity) does determine if the two components appear as bright or dark structures. In the second scenario, when $\alpha=0$ (in which case the value of $\kappa$ becomes irrelevant), system (\ref{helicoidal SOC}) degenerates into the standard Manakov system. Using the method described in this work, the resulting solutions are typical non-oscillating dark-bright soliton pairs.

\subsection{Spin oscillations}

Eq.~(\ref{helicoidal SOC}) is integrable in the absence of the Zeeman
splitting~\cite{Kartashov2017}, and, consequently, it ought to possess an
infinite number of conservation laws that fundamentally constrain the
system's dynamics, albeit these laws seemingly remain unidentified to date.
Here, we delve into the examination of various physical quantities
associated with the wave function of the helicoidal SO-coupled BEC system,
including the particle density and spin density, also scrutinizing the
influence of parameters such as SOC strength $\alpha $ and helical pitch $%
\kappa $ on these quantities. The particle density $n(x,t)$ and spin density
$\boldsymbol{f}(x,t)$ of the system are defined as
\begin{equation}
\begin{aligned}
n(x,t)=&\mathbf{\Psi}^{\dag}\cdot\mathbf{\Psi}-\mathbf{\Psi}[0]^{\dag}\cdot%
\mathbf{\Psi}[0],\\
\boldsymbol{f}(x,t)=&\mathbf{\Psi}^{\dag}\cdot\boldsymbol{\sigma}\cdot%
\mathbf{\Psi}, \end{aligned}  \label{physical quantities}
\end{equation}%
where $\mathbf{\Psi }[0]$ is the initial wave vector given by the seed
solutions~(\ref{plane wave-zero}) and gauge transformation~(\ref{trans1}), $%
\boldsymbol{\sigma }=(\sigma _{1},\sigma _{2},\sigma _{3})$ are the Pauli
matrices and spin density $\boldsymbol{f}(x,t)=(f_{x},f_{y},f_{z})$. In Fig.~%
\ref{fig30}, we present the physical properties of the wave function of
BSSs. Due to the fact that the parameters satisfy the symmetry
condition~(\ref{symmetric stripe}), it is observed that the soliton stripes are now symmetric, as shown in Fig.~\ref%
{fig30}$(a)$. From Eq.~(\ref{physical quantities}), it is also seen that the
particle density distribution $n(x,t)$ takes the form of a soliton, as seen
in Fig.~\ref{fig30}$(b)$. The spin density distribution $(f_{x},f_{y},f_{z})$
exhibits some noteworthy features: with the SOC strength $\alpha \neq 0$
while helical pitch $\kappa =0$, i.e., when SOC is spatially uniform in the
system, the spin density distributions in all three directions exhibit
oscillations near $x=0$ and then tend towards a constant distribution at $%
x\rightarrow \pm \infty $. Among these, $f_{x}$ and $f_{z}$ are even
functions, while $f_{y}$ is an odd one, as shown in Fig.~\ref{fig30}$(c)$.

\begin{figure}[ht]
\includegraphics[scale=0.4]{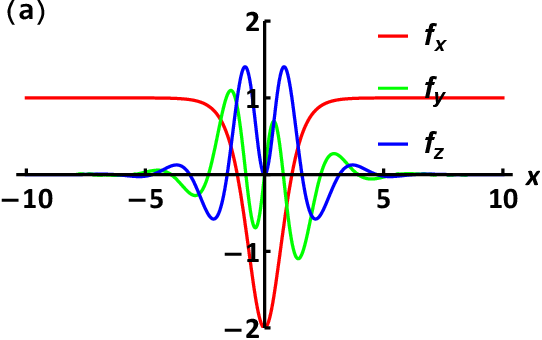}\hspace{3mm} %
\includegraphics[scale=0.4]{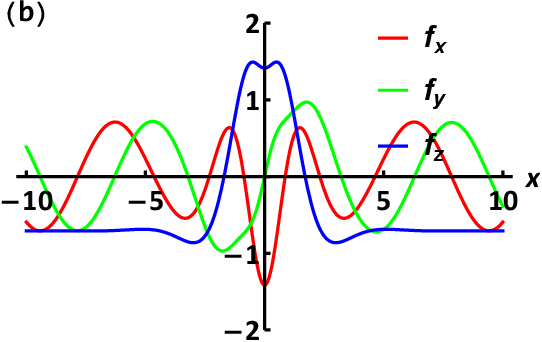}\newline
\vspace{-3mm}
\caption{The role of $(a)$ SOC strength $\protect\alpha $ and $(b)$ helicity
pitch $\protect\kappa $ in modulating the spin density with $\protect\alpha %
=1$ and $\protect\kappa =0$ in $(a)$ and $\protect\alpha =0.5$ and $\protect%
\kappa =0.5$ in $(b)$. The other parameters are the same as Fig.~\protect\ref%
{fig30}.}
\label{fig31}
\end{figure}

Additionally, SOC strength $\alpha $ and helicity pitch $\kappa $ have a
significant effect on the distribution of spin density, as shown in Fig.~\ref%
{fig31}. We first increase $\alpha $ while keeping $\kappa =0$. Then, it is
observed that the spin density distribution oscillates more intensely near $%
x=0$, although it eventually tends towards a constant value. However, for $%
\kappa \neq 0$, i.e., when SOC is non-uniform in the system, spin densities $%
f_{x}$ and $f_{y}$ maintain periodic oscillations and do not tend towards
constant values, as shown in Fig.~\ref{fig31}$(b)$. This is significantly
different from the spin-density distributions in the systems with uniform
SOC or without SOC. We also find that the increase of the helicity pitch $%
\kappa $ leads to increase of the oscillation frequency of the spin density.

\subsection{The composite form of the beating stripe soliton (BSS)}

When all parameters $l_{j}$ $(j=1,2,3)$ are different from zero, solutions~(%
\ref{dark-bright soliton}) characterize the general form of a composite BSS,
with both components featerring a breather and a BSS. The total density of
the components, as a whole, exhibits characteristics akin to both a breather
and a soliton.

\begin{figure*}
\includegraphics[scale=0.35]{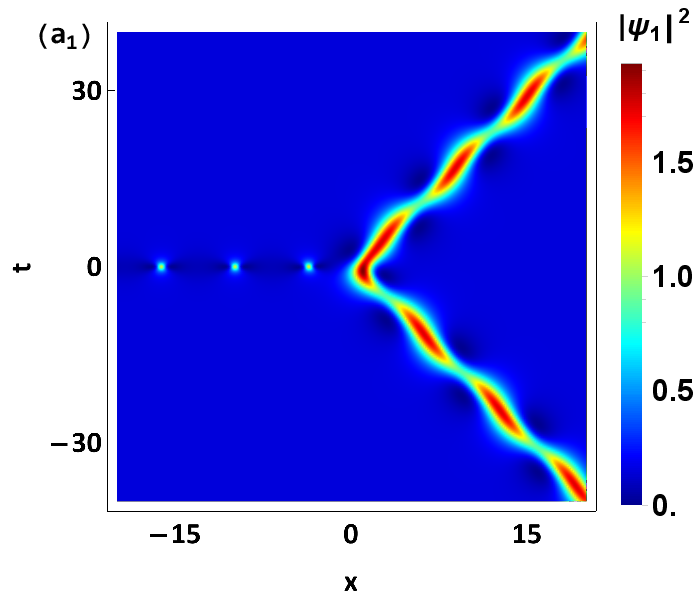}\hspace{3mm}
\includegraphics[scale=0.35]{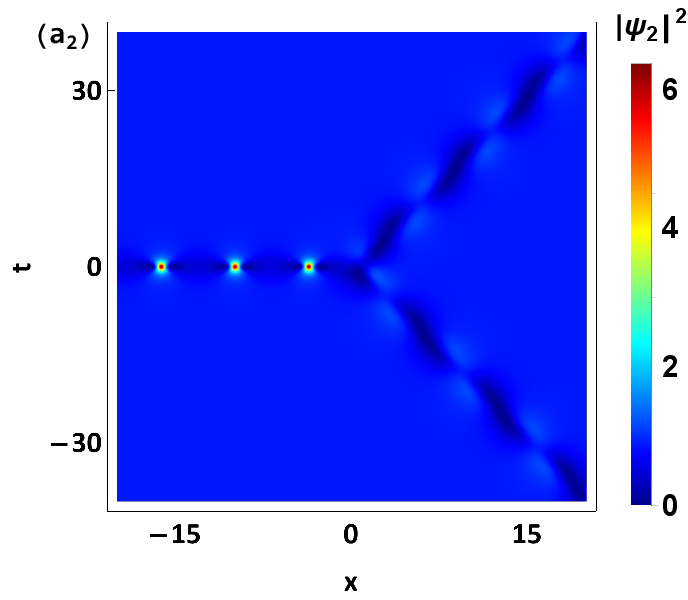} \hspace{3mm}
\includegraphics[scale=0.35]{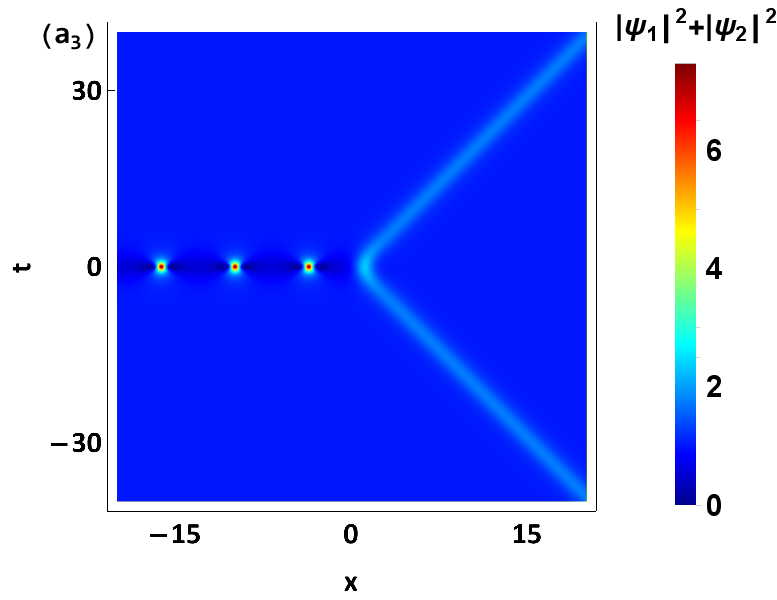} \hspace{3mm}\newline
\includegraphics[scale=0.35]{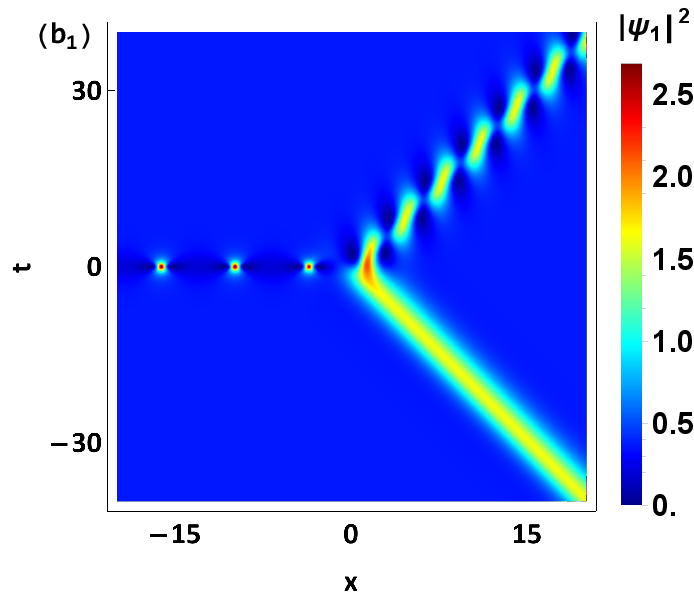}\hspace{3mm}
\includegraphics[scale=0.35]{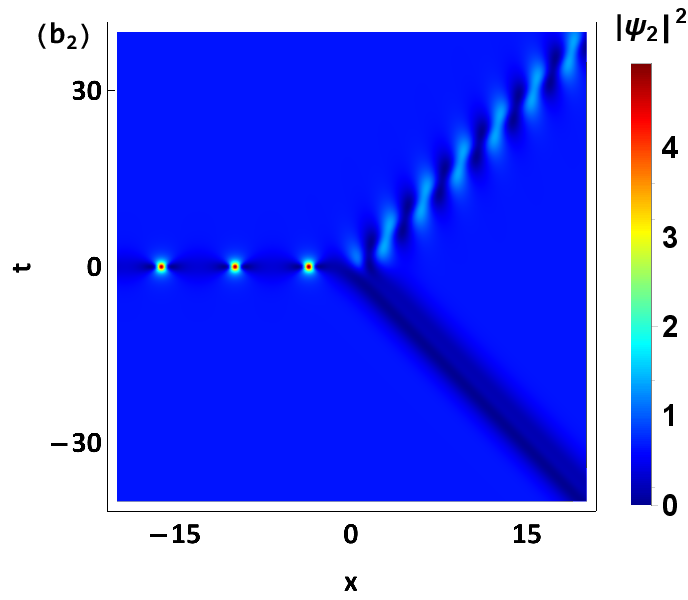} \hspace{3mm}
\includegraphics[scale=0.35]{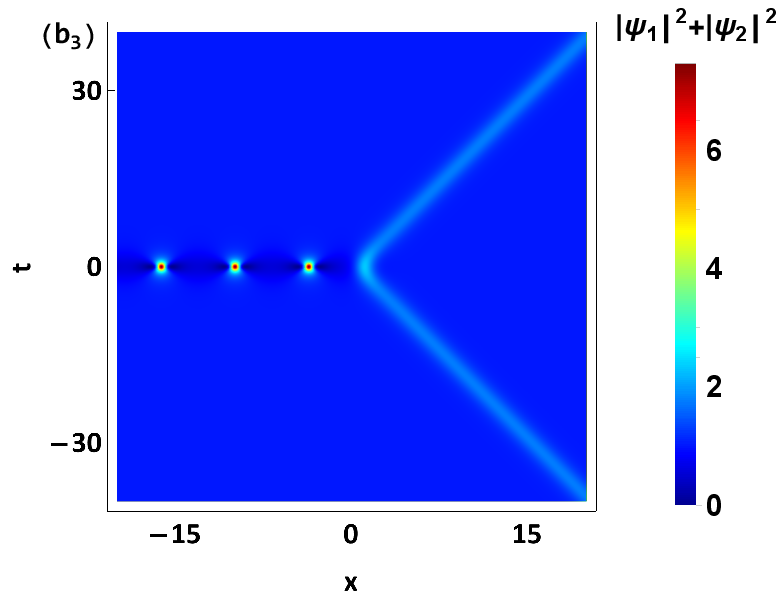} \hspace{3mm}\newline
\includegraphics[scale=0.35]{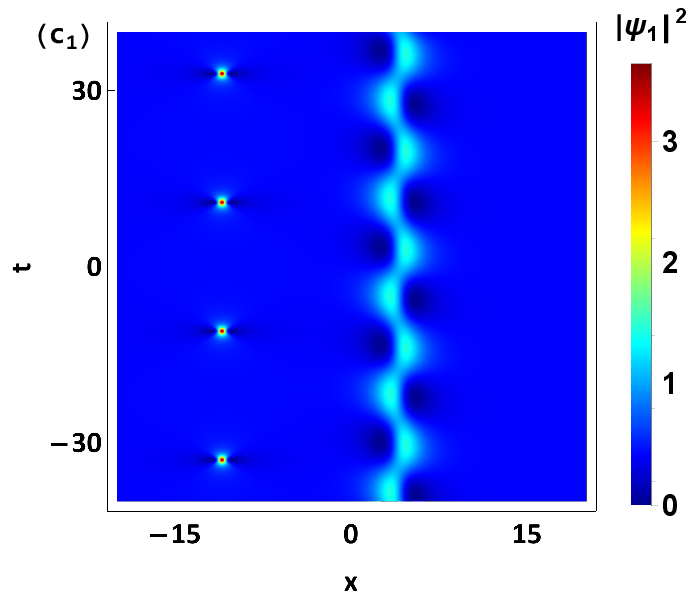}\hspace{3mm}
\includegraphics[scale=0.35]{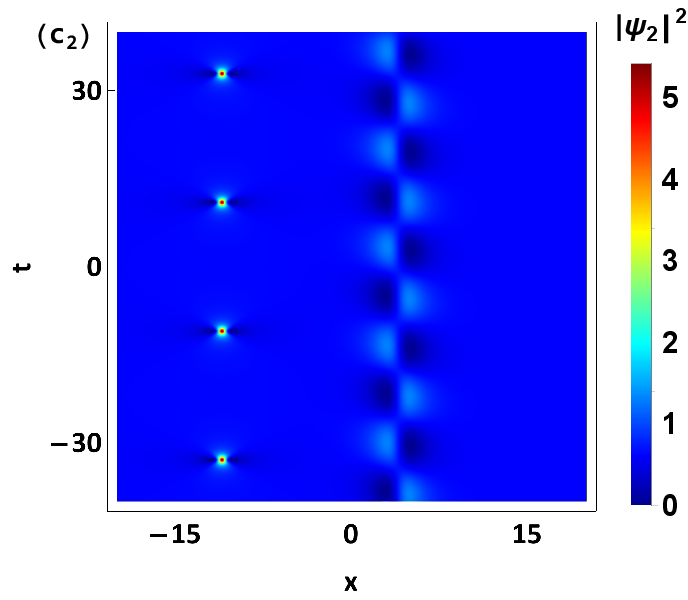} \hspace{3mm}
\includegraphics[scale=0.35]{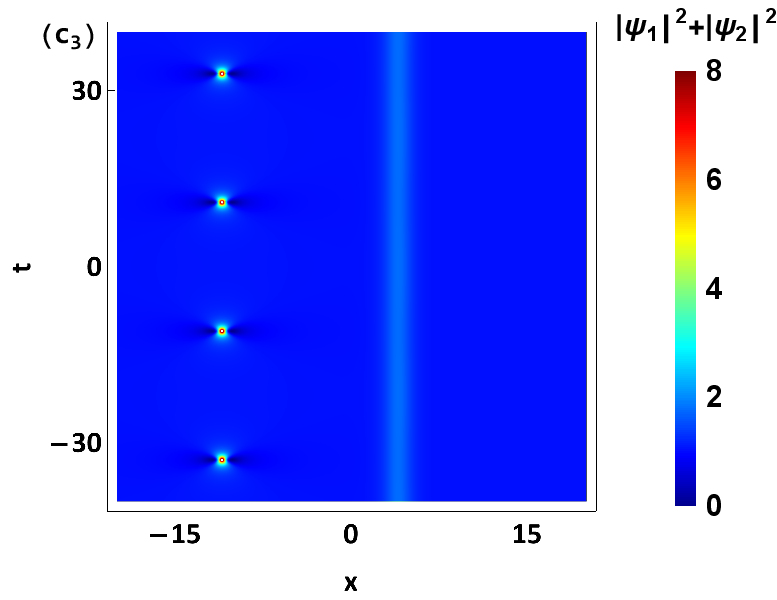} \hspace{3mm}\newline
\vspace{-3mm}
\caption{\label{fig4}Density distributions of various composite beating stripe solitons
given by solutios~(\protect\ref{dark-bright soliton}) with parameters $(a)$ $%
\protect\lambda _{1}=\frac{\protect\sqrt{3}}{2}i$, $\protect\alpha =\protect%
\kappa =0.01$, $l_{1}=l_{2}=l_{3}=1$; $(b)$ $\protect\lambda _{1}=\frac{%
\protect\sqrt{3}}{2}i$, $\protect\alpha =0.48$, $k_{\text{m}}=0.5$, $%
l_{1}=l_{2}=l_{3}=1$; $(c)$ $\protect\lambda _{1}=1.01i$, $\protect\alpha %
=0.05$, $\protect\kappa =0.01$, $l_{1}=l_{3}=1$, $l_{2}=20$. The other
parameters are $s=1$, $a=1$, $k_{1}=0$.}
\end{figure*}

Figure~\ref{fig4} depicts several varieties of such solitons. Analysis
reveals that, at $sa^{2}-\lambda _{1I}^{2}>0$, the breather propagates
periodically parallel to the $x$-axis, until it encounters a BSS, resulting
in a transformed BSS that alters its direction of propagation, as shown in
Fig.~\ref{fig4}$(a)$. As depicted and analyzed in Fig.~\ref{fig2}$(a)$, when
the parameters fulfill the condition $k_{t}/k_{x}=-\mu _{1R}$, the BSS
degenerates into an oscillating bright-dark soliton pair. Subsequently, we
can obtain a composite BSS, which is a union of a breather and the
oscillating bright-dark soliton pair, as illustrated in Fig.~\ref{fig4}$(b)$%
. Notably, the oscillating bright-dark soliton pair here displays a
pronounced oscillation period and, due to the swift decay of the amplitude
decay to the background plane, it assumes an $M$-shaped form. Despite this,
the total density of the two components maintains its shape, consisting of a
breather and a soliton. Additionally, it is evident that, upon the collision
with the breather, this $M$-shaped bright-dark soliton pair evolves into a
conventional BSS. Here, the period of the stripe solitons can be controlled
by adjusting SOC, thereby changing the number of stripes to form $M$-shaped
solitons or other forms of stripe solitons. Moreover, in the scenario with $%
sa^{2}-\lambda _{1I}^{2}<0$, the breather and the beating soliton keep their
parallel alignment, and it is possible to fine-tune the distance between
them by adjusting parameter $l_{j}$. A characteristic example showcasing
this phenomenon is depicted in Fig.~\ref{fig4}$(c)$. Essentially, the ratio $%
|l_{2}|/|l_{1}|$ plays the role of a modulator, allowing one to adjust the
horizontal coordinate of the intersection point between the breather and
BSS, whereas $|l_{3}|/|l_{1}|$ serves to alter the vertical position of this
point. Accordingly, the positions of the interaction between the soliton and
breather can be controlled too. This principle holds true for a diverse
array of composite BSSs.

\begin{figure}[ht]
\includegraphics[scale=0.25]{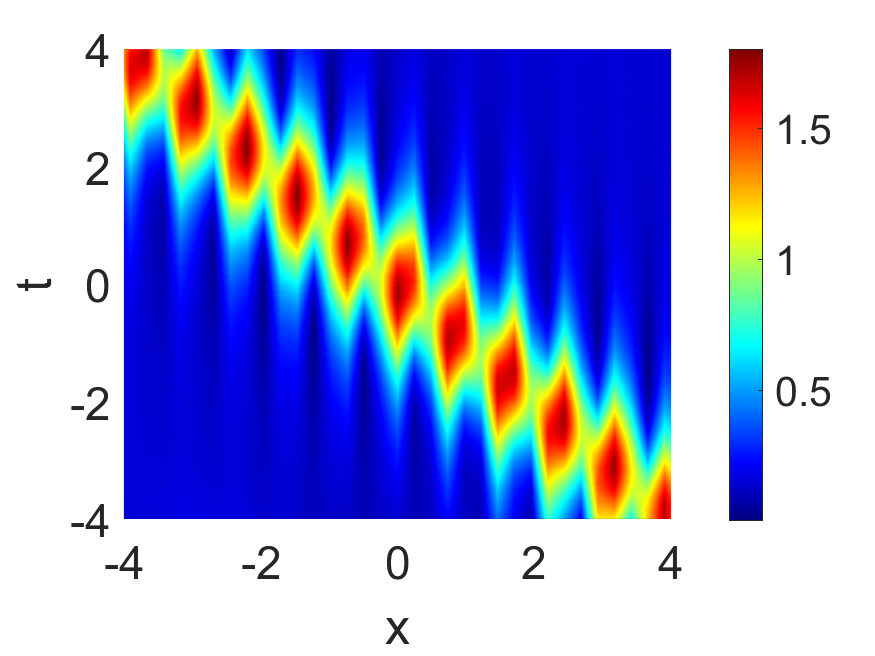}\hspace{3mm} %
\includegraphics[scale=0.25]{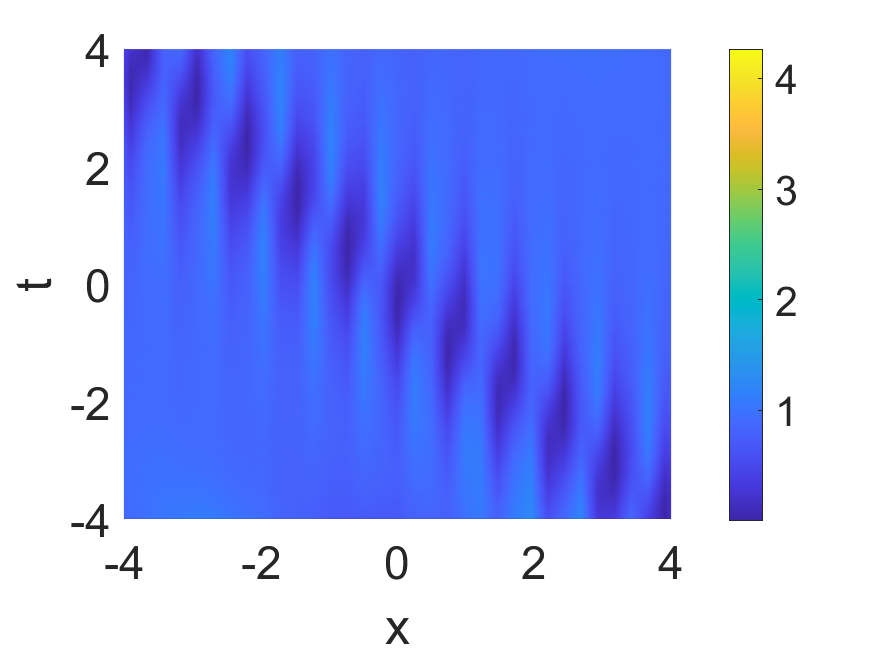}\newline
\includegraphics[scale=0.25]{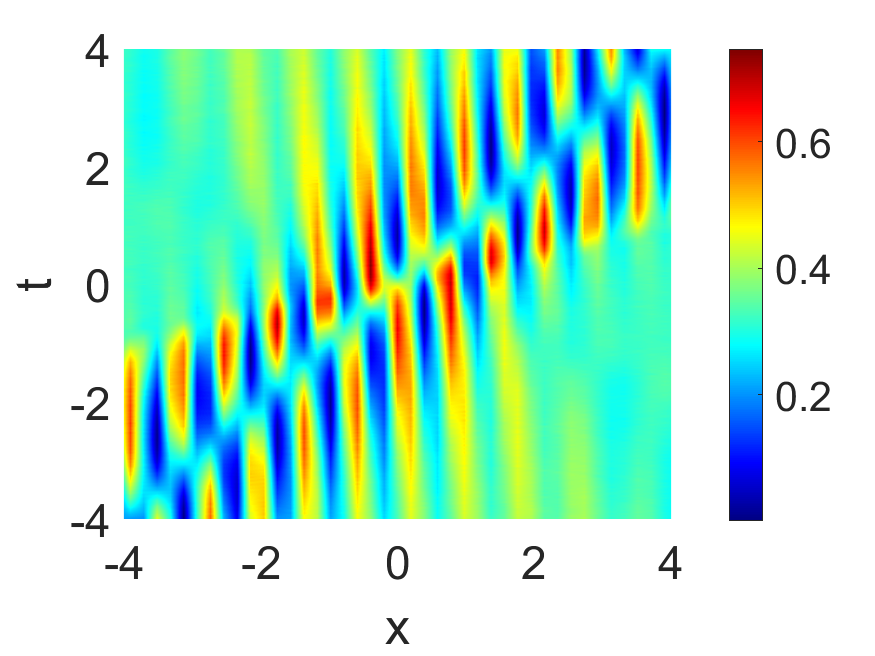}\hspace{3mm} %
\includegraphics[scale=0.25]{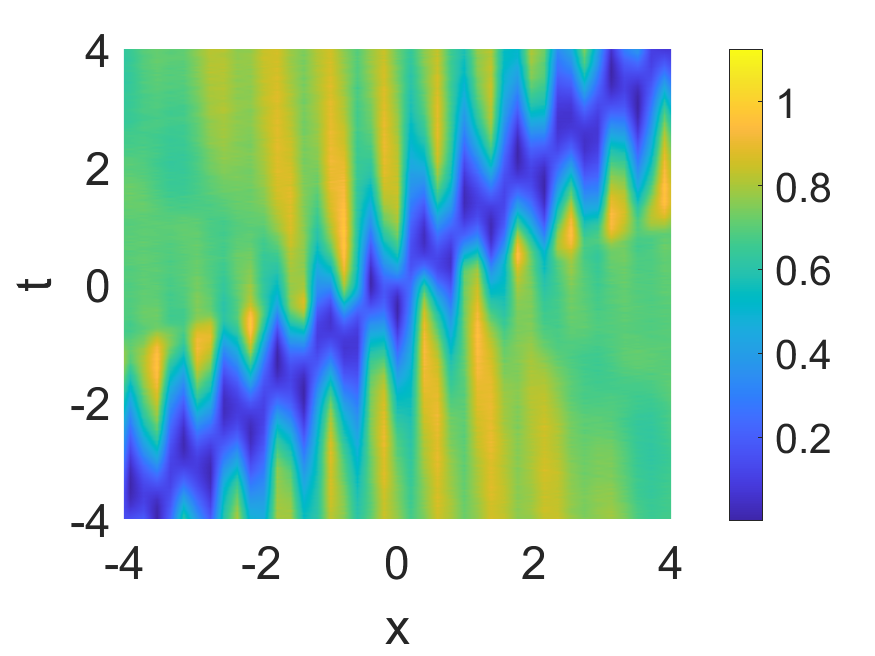}\newline
\includegraphics[scale=0.25]{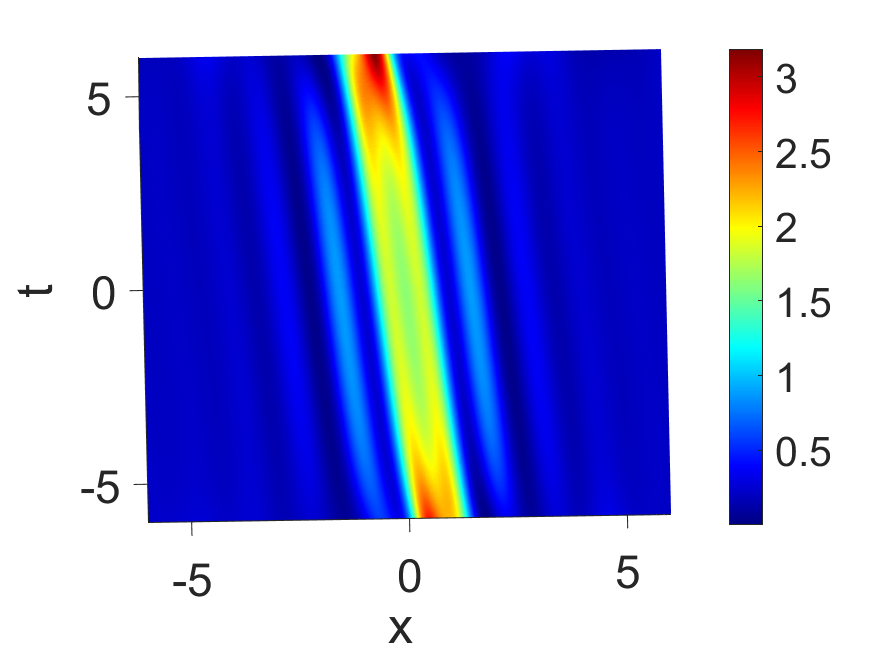}\hspace{3mm} %
\includegraphics[scale=0.25]{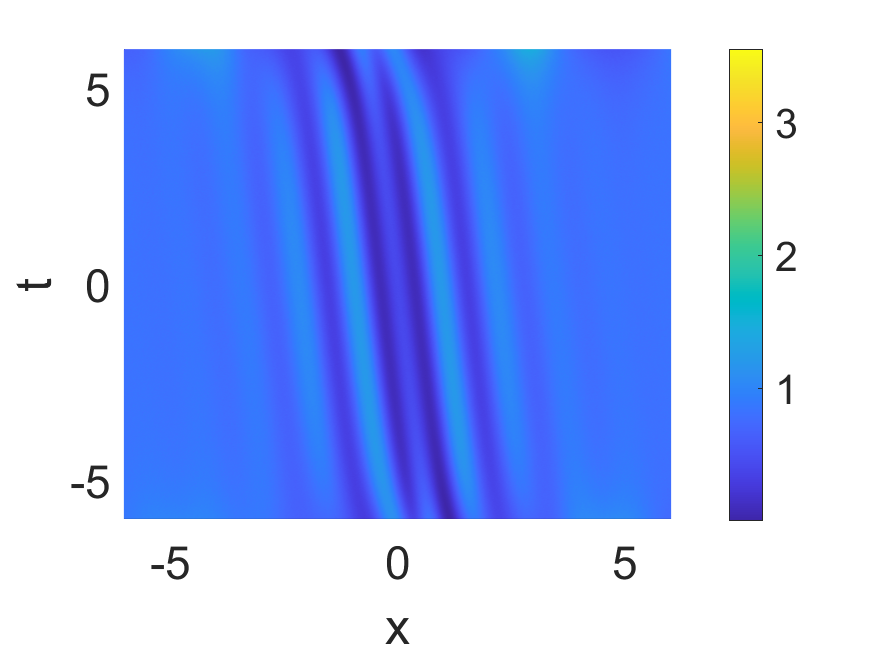}\newline
\vspace{-3mm}
\caption{The results of the numerical simulations of the evolution of BSSs under the action of 2\% noise.
The initial state is taken as per the analytical solutions~(\ref{Stripe dark-bright}). The top row corresponds to Fig.~\ref{fig1}$(a)$,
the middle row corresponds to Fig.~\ref{fig1}$(b)$, and the bottom row corresponds to Fig.~\ref{fig2}$(a)$.}
\label{fig32}
\end{figure}

To complete the discussion of BSSs, we conduct numerical simulations to study their stability. By using the split-step Fourier method and the fourth-order Runge-Kutta method, with the analytical solutions~(\ref{Stripe dark-bright}) as the initial solution, the evolution of BSSs under a 2\% noise perturbation is obtained. The results are shown in Fig.~\ref{fig32}. It can be seen that the basic BSSs in the attractive and repulsive regimes, as well as the non-oscillating bright-dark soliton pairs, can remain basically stable under certain perturbations.

Lastly, it is relevant to compare the present results with previous studies of stripe solitons in the presence of the helicoidal SOC-BEC. The existence and stability of freely moving bright stripe solitons, induced by the linear superposition of bright-bright solitons and the stripe state, were addressed in Ref. \cite{Kartashov2017} (see Eq. (5) of Ref. \cite{Kartashov2017}). In contrast, the present paper primarily examines solitons in the helicoidal SOC-BEC that feature both stripe and beating states, formed by the superposition of dark-bright solitons on the non-zero background.

\section{Nonlinear superposition of the beating stripe solitons (BSSs)}

\subsection{The nonlinear superposition of the fundamental BSSs}

\begin{figure*}
\includegraphics[scale=0.4]{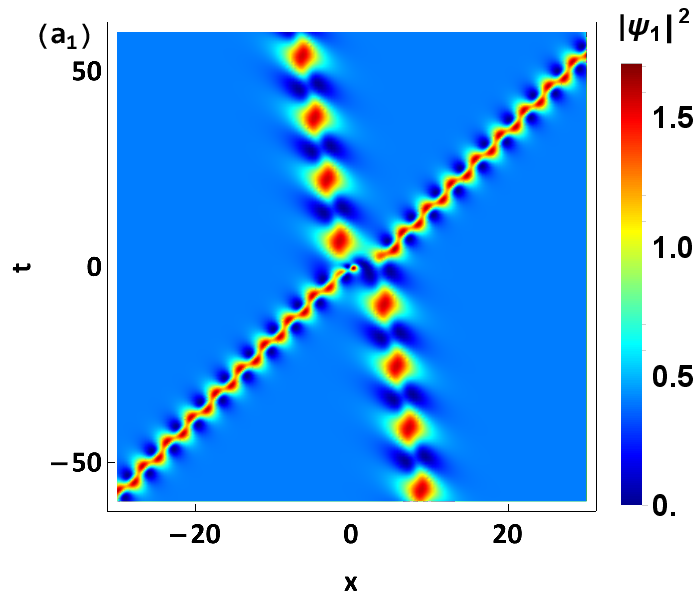}\hspace{3mm}
\includegraphics[scale=0.4]{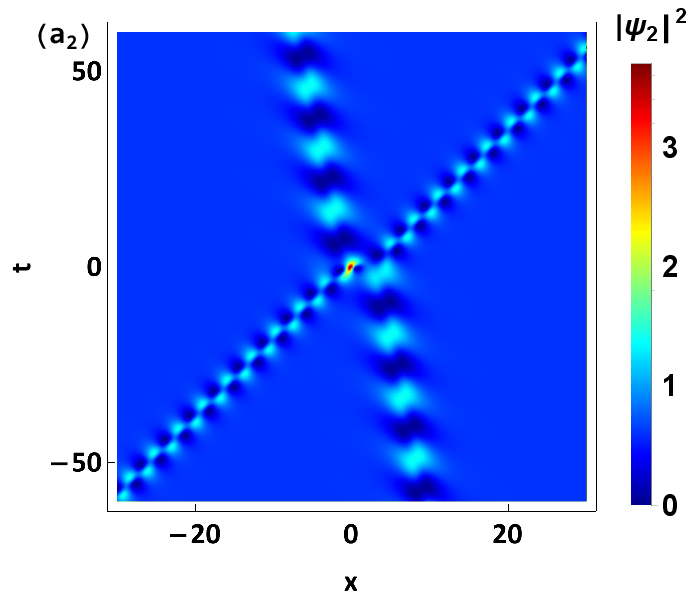} \hspace{3mm}
\includegraphics[scale=0.4]{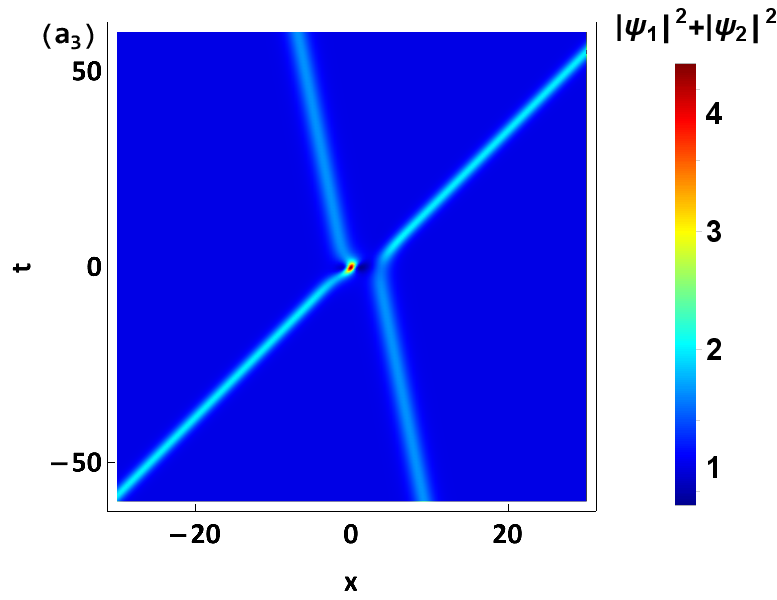} \hspace{3mm}\newline
\includegraphics[scale=0.4]{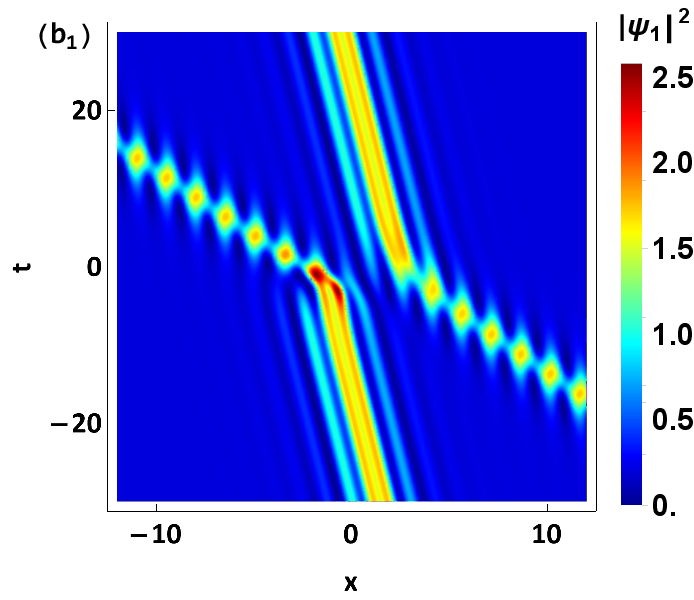}\hspace{3mm}
\includegraphics[scale=0.4]{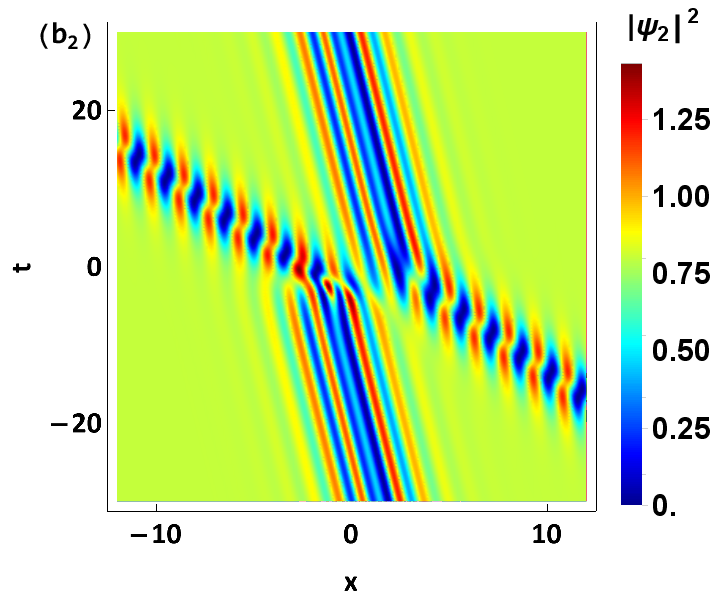} \hspace{3mm}
\includegraphics[scale=0.4]{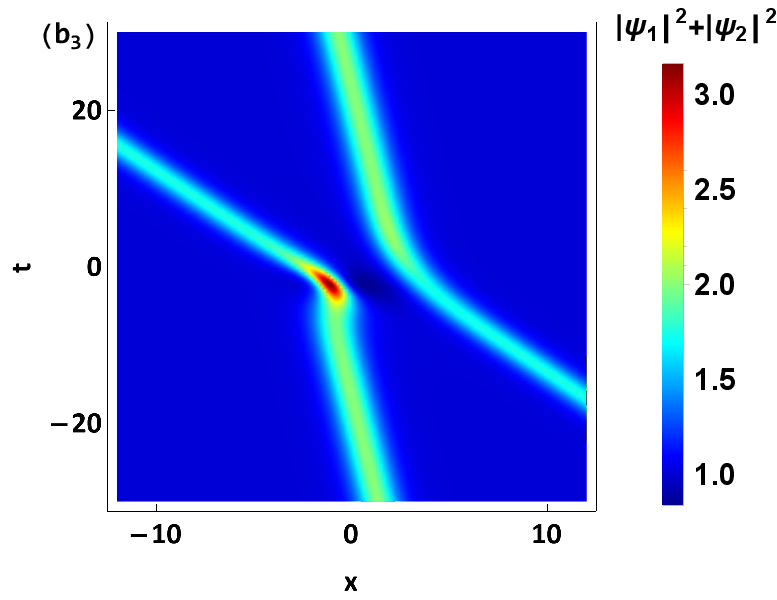} \hspace{3mm}\newline
\includegraphics[scale=0.4]{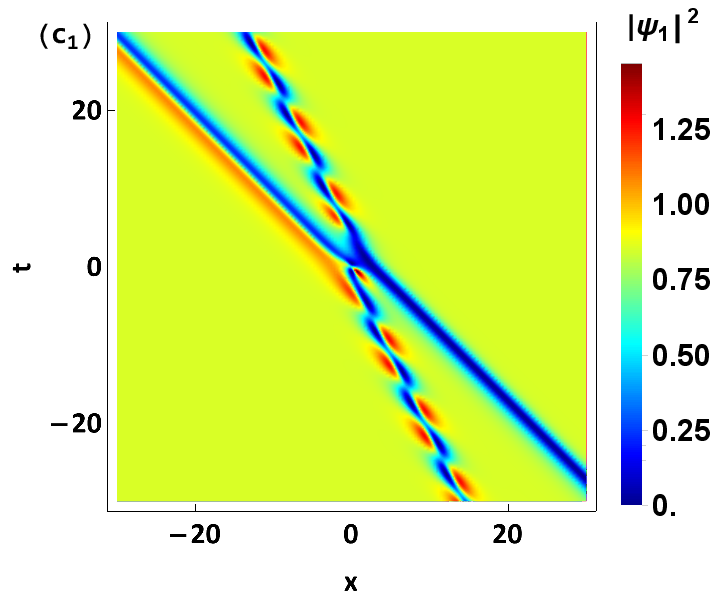}\hspace{3mm}
\includegraphics[scale=0.4]{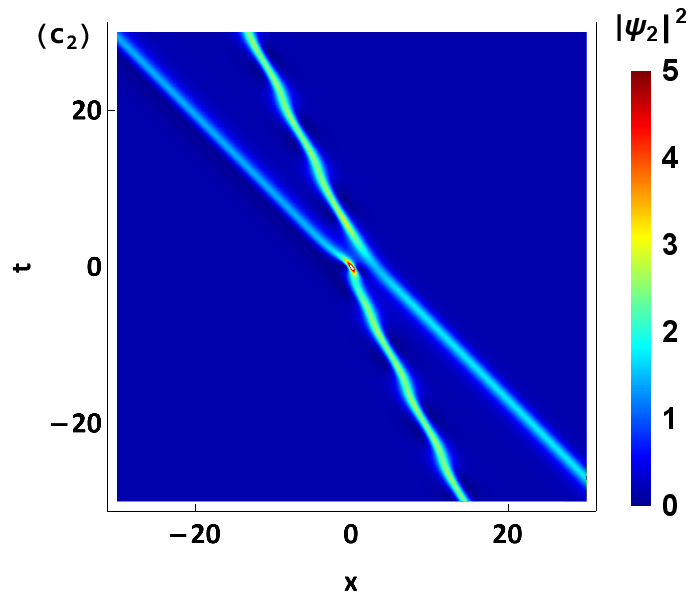} \hspace{3mm}
\includegraphics[scale=0.4]{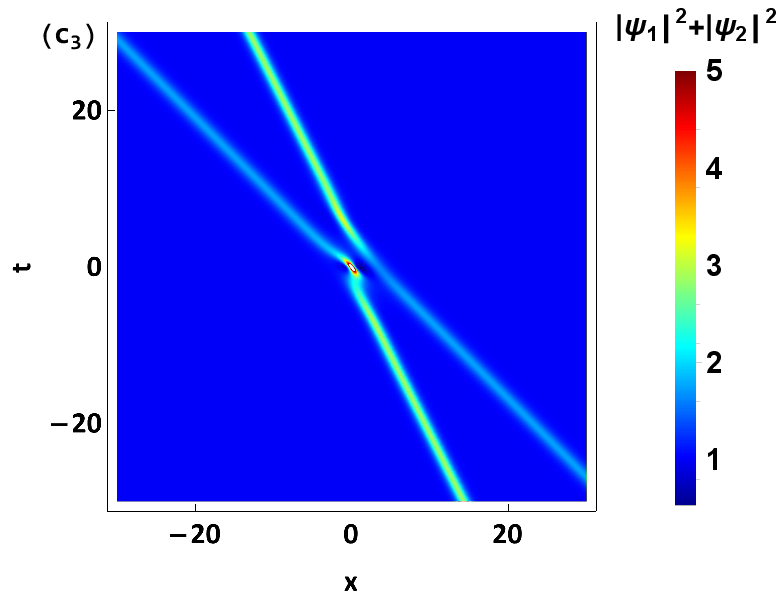} \hspace{3mm}\newline
\vspace{-3mm}
\caption{\label{fig5}(a) The interaction between two beating stripe solitons with $%
k_{1}=\frac{1}{2}$, $\protect\lambda _{1}=-\frac{1}{4}+\frac{4}{5}i$, $%
\protect\lambda _{2}=-\frac{1}{4}+i$, $\protect\alpha =0.5$, and $\protect%
\kappa =0.1$; (b) The interaction between a beating stripe soliton and a
vibrating dark-bright soliton pair with $k_{1}=-0.1$, $\protect\lambda %
_{1}=0.05-i$, $\protect\lambda _{2}=0.05+\frac{\protect\sqrt{3}}{2}i$, $%
\protect\alpha =2$, and $\protect\kappa =1.5$; (c) The interaction between
a beating stripe soliton and non-vibrating dark-bright soliton pair with $%
k_{1}=-0.5$, $\protect\lambda _{1}=0.25+\frac{\protect\sqrt{3}}{2}i$, $%
\protect\lambda _{2}=0.25+1.05i$, $\protect\alpha =-\protect\kappa =\frac{%
\protect\sqrt{2}}{8}$. The other parameters are $s=1$, $a=1$, $%
l_{j1}=l_{j3}=1$, and $l_{j2}=0$ $(j=1,2)$.}
\end{figure*}

In this subsection, we conduct address the nonlinear superposition of BSSs
through the $N$-th-order solutions derived in Appendix B, and analyze the
interaction between the two solitons. Firstly, we consider the attraction
regime with $s=1$. In this particular scenario, the velocities of the two
solitons can vary, depending on the choice of spectral parameters $\lambda $%
. Specifically, for $sa^{2}-\lambda _{I}^{2}\geq 0$, the velocity\ is $%
V_{g}=k_{1}-\sqrt{sa^{2}-\lambda _{I}^{2}}$, whereas for $sa^{2}-\lambda
_{I}^{2}\leq 0$ it is $V_{g}=k_{1}$. To study the interaction between
different solitons, we manipulate one soliton to exhibit both beating and
striped states simultaneously, while modifying the morphology of the other
soliton, including the BSS, the stripe soliton lacking a beating pattern,
and the single soliton without any beating or striping. Illustrative
examples of head-on collisions involving various types of BSSs are displayed
in Fig.~\ref{fig5}. It is evident that, although solitons
maintain their beating and/or stripe states after interaction,
there are subtle changes in their shapes, in addition to the interaction-induced phase shift.

\begin{figure*}
\includegraphics[scale=0.4]{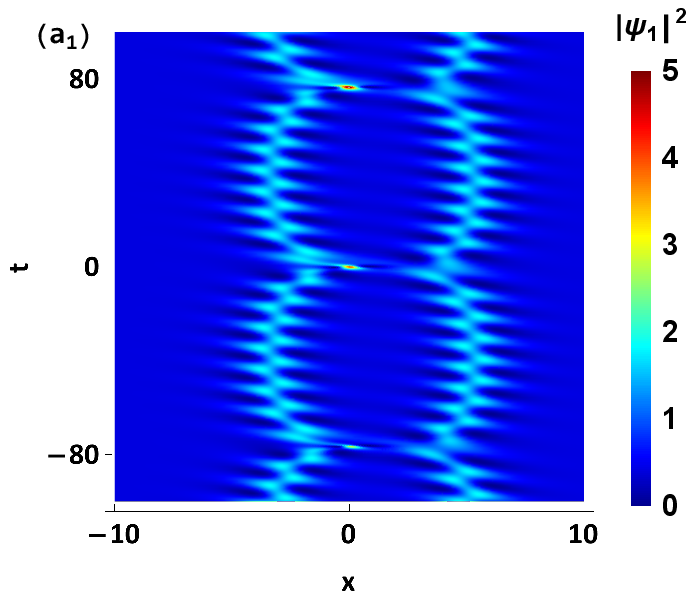}\hspace{3mm}
\includegraphics[scale=0.4]{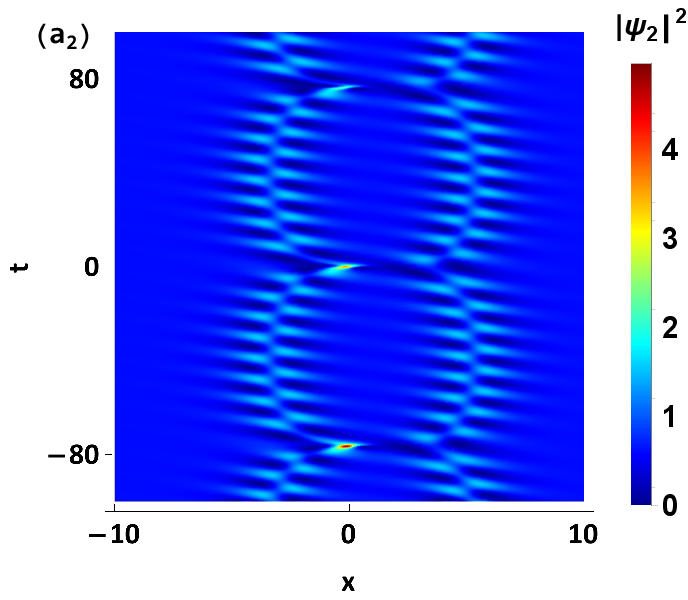} \hspace{3mm}
\includegraphics[scale=0.4]{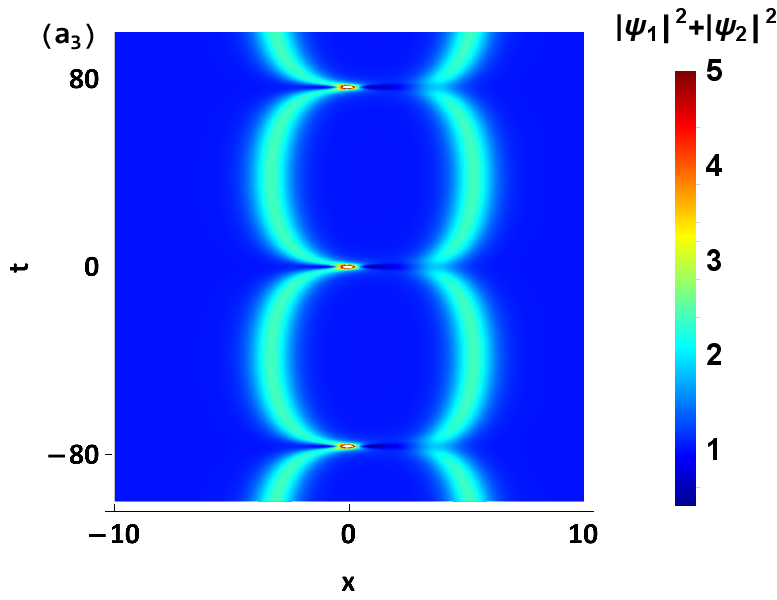} \hspace{3mm}\newline
\includegraphics[scale=0.4]{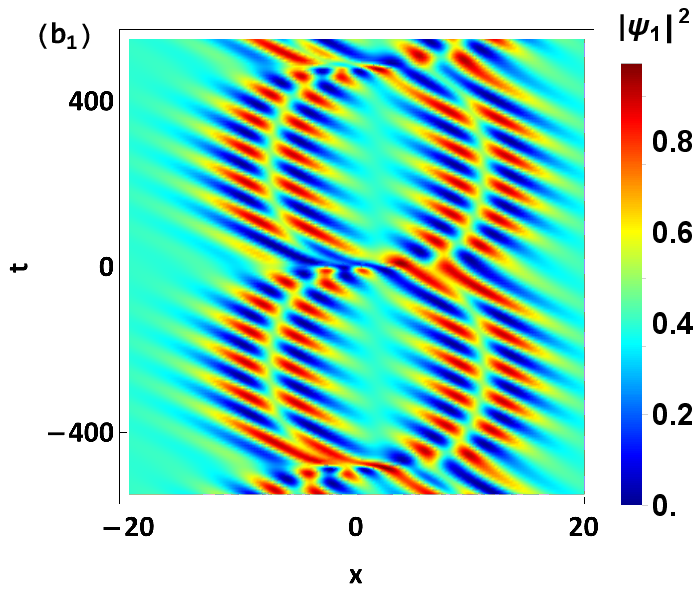}\hspace{3mm}
\includegraphics[scale=0.4]{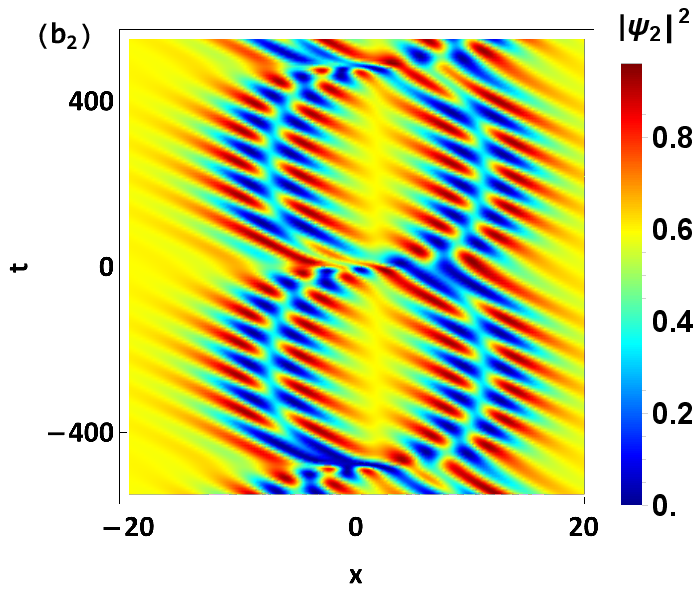} \hspace{3mm}
\includegraphics[scale=0.4]{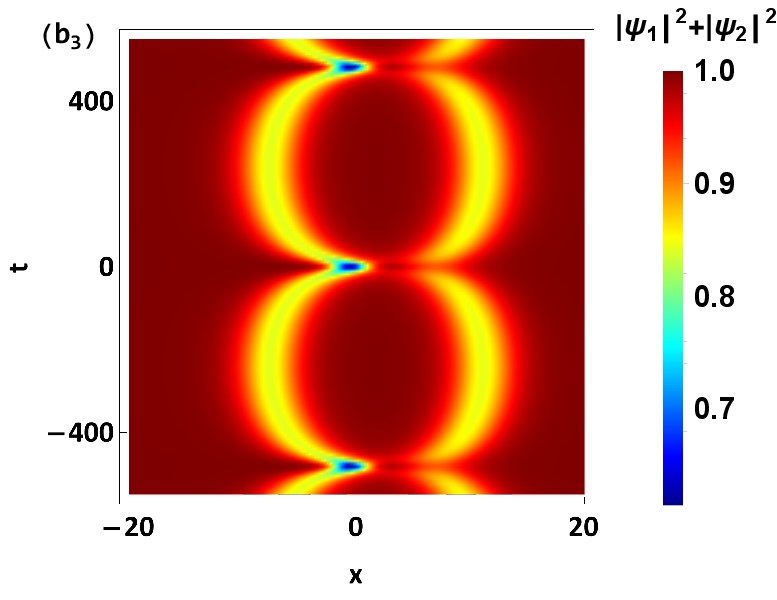} \hspace{3mm}\newline
\includegraphics[scale=0.4]{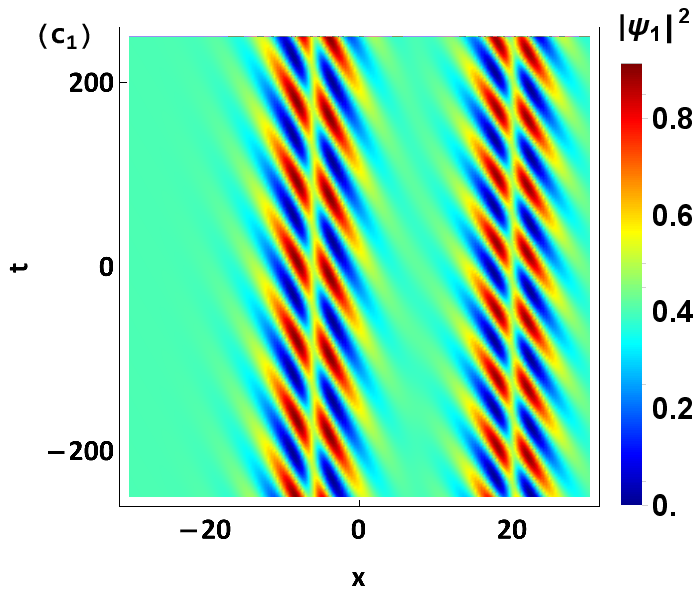}\hspace{3mm}
\includegraphics[scale=0.4]{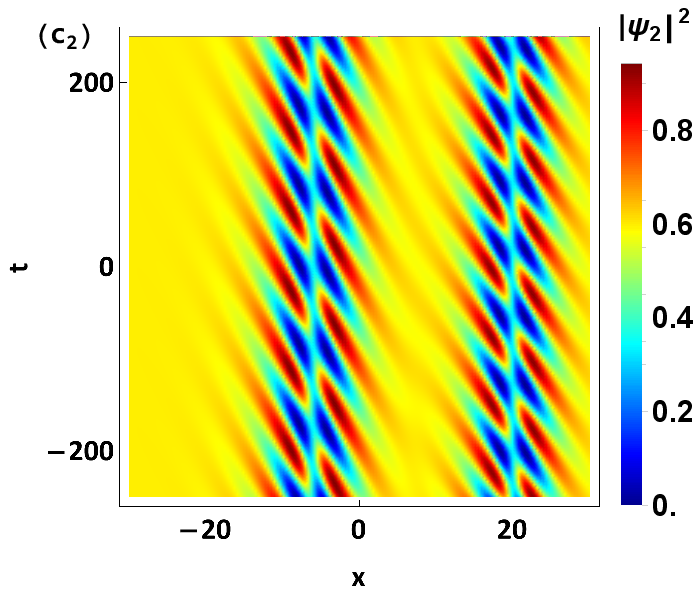} \hspace{3mm}
\includegraphics[scale=0.4]{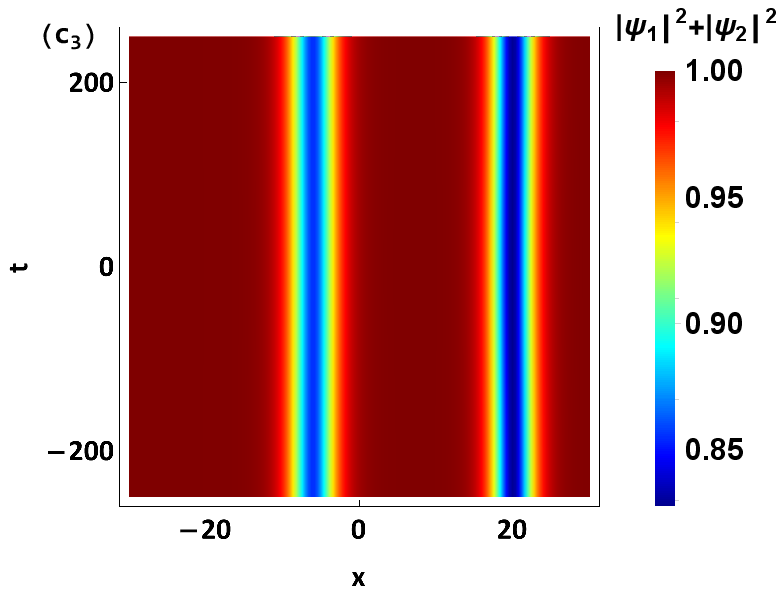} \hspace{3mm}\newline
\vspace{-3mm}
\caption{\label{fig6}Bound and parallel states of two beating stripe solitons with
parameters (a) $s=1$, $\protect\lambda _{1}=1.01i$, $\protect\lambda %
_{2}=1.02i$; (b) $s=-1$, $\protect\lambda _{1}=-i$, $\protect\lambda %
_{2}=-1.12i$; (c) the same parameters as in (b), except for $l_{11}=100$.
The other parameters are $a=1$, $k_{1}=0$, $\protect\alpha =0.5$, $\protect%
\kappa =0.1$, $l_{j1}=l_{j3}=1$, and $l_{j2}=0$ $(j=1,2)$.}
\end{figure*}

In addition to the head-on collision that occurs when solitons travel at
different speeds, bound solitons can be formed when two solitons possess
identical velocities. Under such circumstances, the two BSSs assume a bound
or parallel state, as shown in Fig.~\ref{fig6}. In the attraction regime, we
can take $\lambda _{1,2I}\geq 1$ to obtain two solitons with identical
velocities $V_{g}=k_{1}$; then, the BSS with a bound or parallel state can
emerge. In the case of repulsion, given that the velocity of the solitons
remains constant and equivalent to $k_{1}$, the two solitons invariably form
the bound or parallel state, as illustrated in Fig.~\ref{fig6}. When the
distance between the two solitons is relatively short (adjustable via $l_{1}$%
), a periodic bound state, characterized by alternating attraction and
repulsion can be observed, as shown in Figs.~\ref{fig6} $(a)$ and $(b)$.
Conversely, when the solitons are situated farther apart, they maintain a
parallel arrangement, as shown in Fig.~\ref{fig6} $(c)$. It is worth
pointing out that SOC strength $\alpha $ and helicity pitch $\kappa $, which
have no impact on the velocity of soliton, do not affect their bound or
parallel states either. Instead, they merely modify the solitons' inherent
beating and stripe states. BSSs, which are parallel to
each other with a bright total density in the attraction regime, can also be
obtained by adjusting the distance between the two solitons when they
possess the same velocity.

\subsection{The nonlinear superposition of the composite BSS}

\begin{figure*}
\includegraphics[scale=0.4]{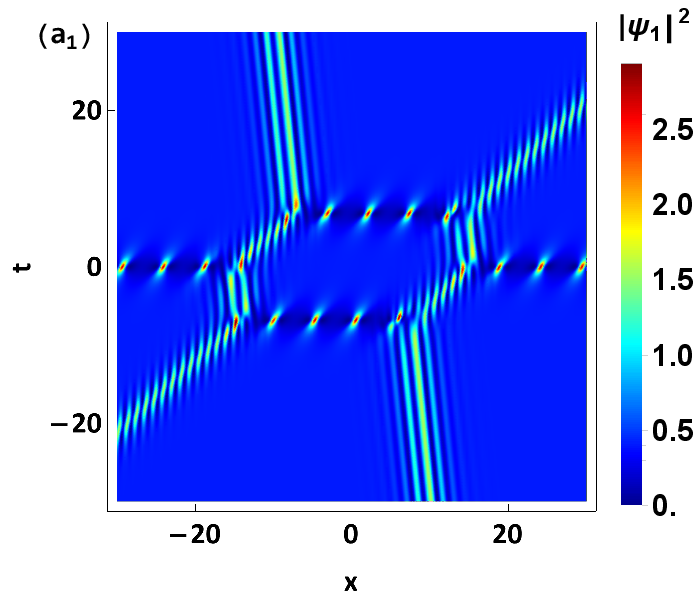}\hspace{3mm}
\includegraphics[scale=0.4]{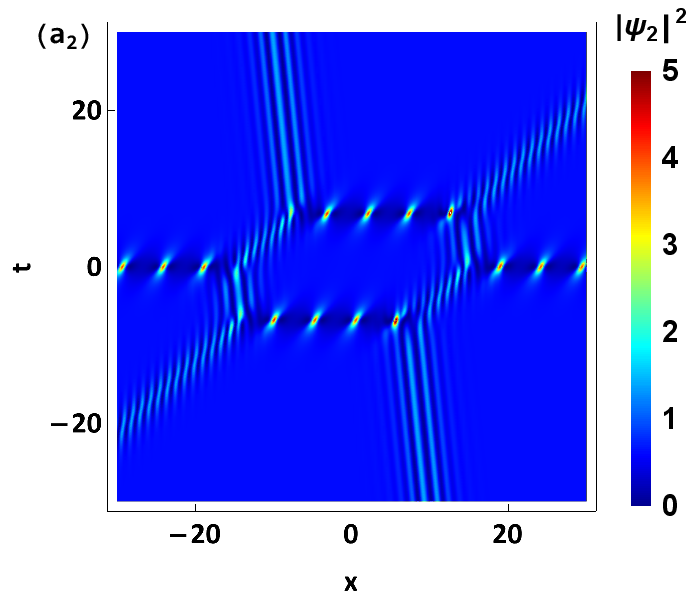} \hspace{3mm}
\includegraphics[scale=0.4]{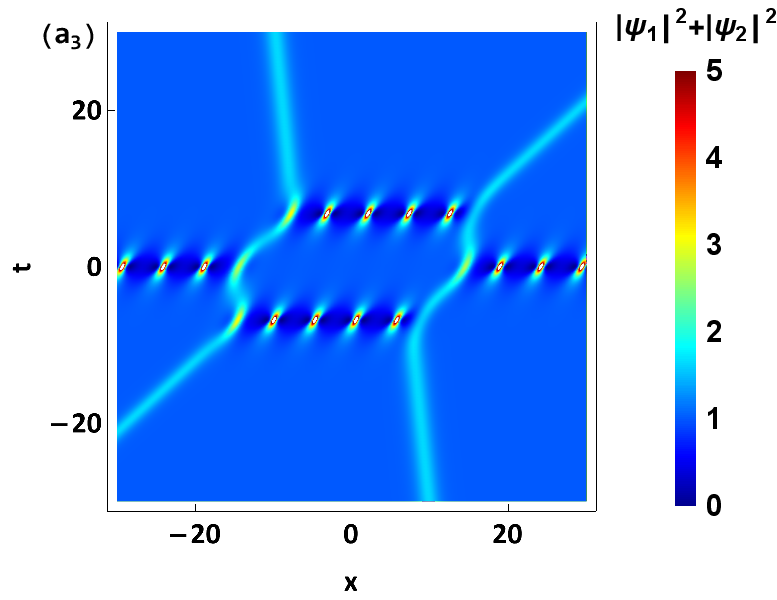} \hspace{3mm}\newline
\includegraphics[scale=0.4]{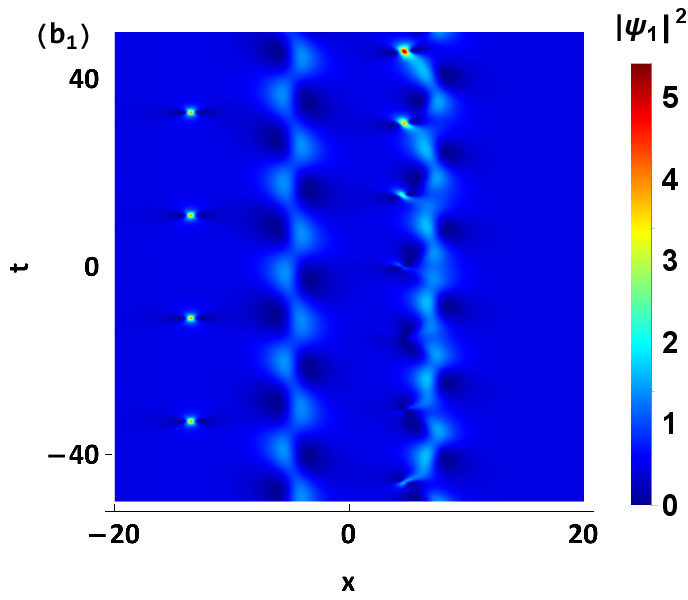}\hspace{3mm}
\includegraphics[scale=0.4]{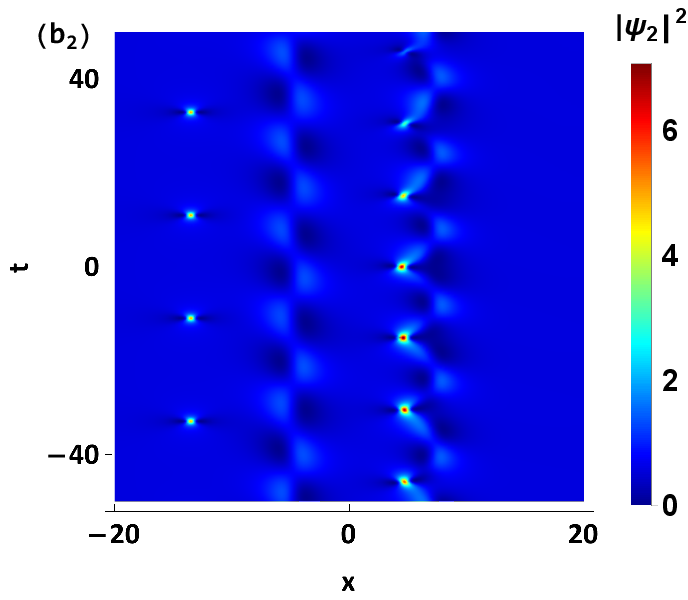} \hspace{3mm}
\includegraphics[scale=0.4]{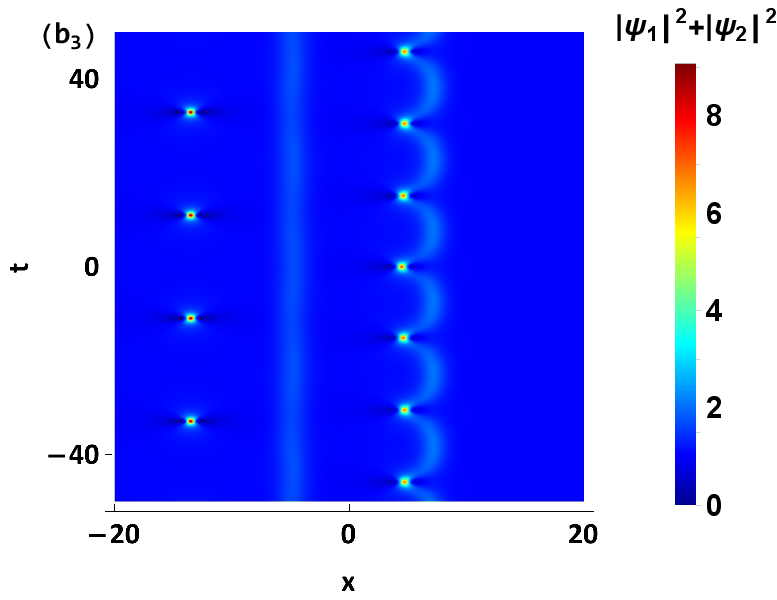} \hspace{3mm}\newline
\includegraphics[scale=0.4]{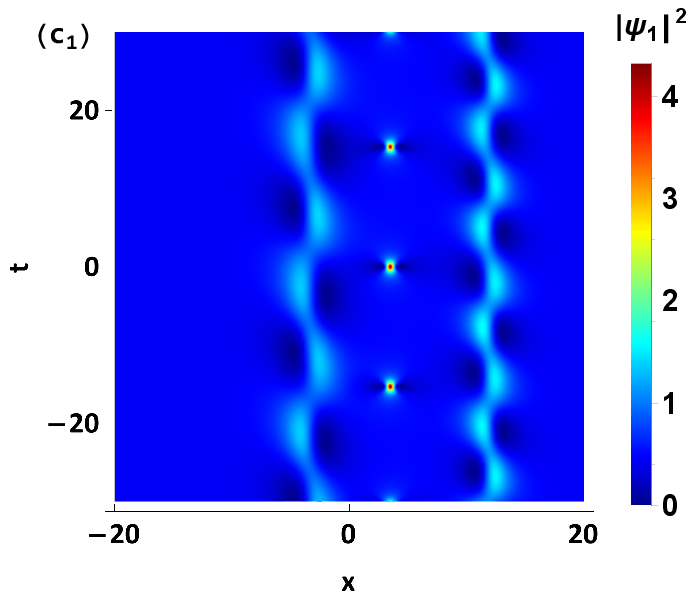}\hspace{3mm}
\includegraphics[scale=0.4]{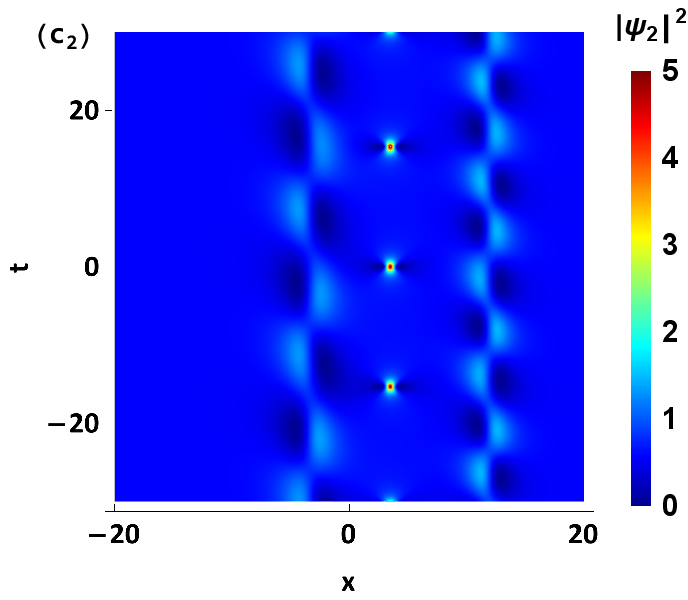} \hspace{3mm}
\includegraphics[scale=0.4]{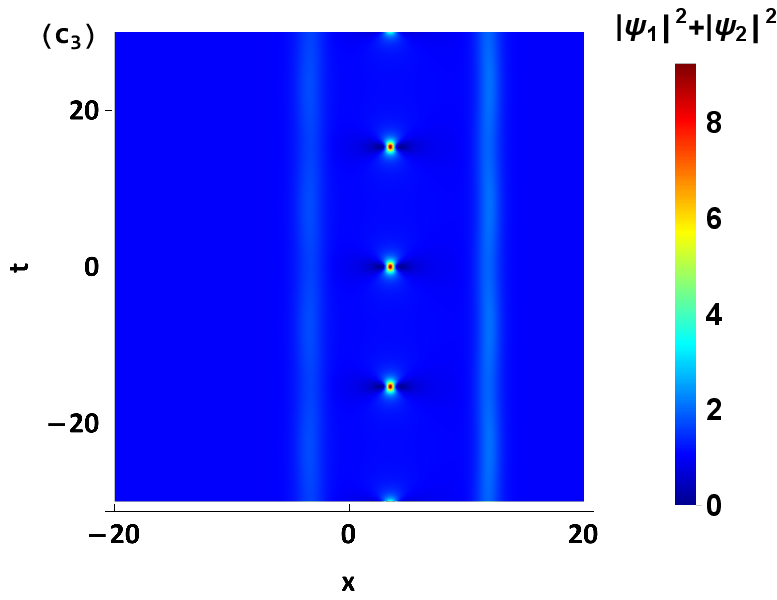} \hspace{3mm}\newline
\vspace{-3mm}
\caption{\label{fig7}(a) Interactions between two $Y$-shaped composite beating stripe
solitons with $k_{1}=0.5$, $\protect\lambda _{1}=-0.25+0.8i$, $\protect%
\lambda _{2}=-0.25-0.801i$, $\protect\alpha =\protect\sqrt{6}$, $\protect%
\kappa =0.5$, $l_{13}=-l_{23}=0.01$. (b) Two parallel composite beating
stripe solitons with $k_{1}=0$, $\protect\lambda _{1}=1.01i$, $\protect%
\lambda _{2}=-1.02i$, $\protect\alpha =0.1$, $\protect\kappa =0.01$, $%
l_{12}=40$;. (c) Mutually parallel a composite beating stripe soliton and a
single beating stripe one with $k_{1}=0$, $\protect\lambda _{1}=-1.02i$, $%
\protect\lambda _{2}=i$, $\protect\alpha =0.1$, $\protect\kappa =0.01$, $%
l_{12}=0.3$, $l_{22}=10000$. The other parameters are $s=1$, $a=1$, and $%
l_{11}=l_{21}=1$.}
\end{figure*}

Next, we delve into a more comprehensive scenario involving the nonlinear
superposition of two composite BSSs, thereby investigating their
interactions. In this context, all coefficients $l_{ij}$ are different from
zero. Initially, we examine a scenario where the solitons intersect rather
than align in parallel, with the directions of the two solitons configured
to be opposite. This setup is illustrated in Fig.~\ref{fig7}$(a)$. In the
attraction regime, one soliton exclusively exhibits a stripe shape, devoid
of a beating state, while the other soliton carries both the beatings and
stripe structure. Notably, the interaction between the two $Y$-shaped
solitons, oriented in diverging directions, generates a hexagon at their
junction, which is composed of (beating) stripe solitons and breathers. The
interaction is quasi-elastic, as both types of solitons and the breather
retain their structures intact post-interaction.

Next, we examine the interactions between two parallel composite solitons in
the attraction regime, as shown in Figs.~\ref{fig7}$(b)$ and $(c)$. When the
imaginary part of the spectral parameter $\lambda $ is $\geq 1$, the two
composite solitons share the same velocity and maintain their parallel
alignment. Additionally, the distance between the soliton and breather can
be adjusted by varying the values of $l_{ij}$. Notably, in the case of Im$%
\left( \lambda \right) =1$, one component degrades, featuring solely BSS
without a breather, as illustrated in Figs.~\ref{fig7}$(c)$. Similar
properties are demonstrated by interactions between other types of composite
solitons.

\section{Conclusions}

In summary, we have investigated the BSSs (beating stripe solitons) in the
two-component BEC under the action of the non-uniform helicoidal spin-orbit
coupling (SOC). By establishing the possibility of its transformation to the
integrable Manakov system, we have constructed fundamental and $N$-th order
soliton solutions for the original BEC system, in the case of both
attractive and repulsive interactions. These solutions describe spatial
and/or temporal periodic breathing oscillations of the density distribution
in each component, which simultaneously possess beating and striped phases,
while the total density does not exhibit oscillations. Solutions for various
moving and static bright/dark/four-petal solitons, along with three
degenerate forms, including bright-dark soliton pairs with or without
oscillations and complementary beating solitons, were obtained. We have
conducted the analysis of the propagation dynamics and established
parametric conditions for the existence of such solitons. Further analysis
revealed that helicoidal SOC significantly affects the structure of such
solitons. Conditions for the generation of symmetric and asymmetric BSSs
were reported too. In the study of physical quantities, such as the particle
and spin densities, we have observed the phenomenon of spin oscillations,
where the spin-density distribution undergoes periodic oscillations, the
stability and frequencies of the oscillations being controlled by the
helicoidal SOC. Additionally, we have obtained a composite soliton
containing both breathers and BSSs, and analyzed its diverse dynamical
behavior. Finally, we studied the nonlinear superposition and interaction
between different BSSs, including head-on collisions between solitons with
different velocities, and bound states formed by solitons with identical
velocities.

In the experiment, a laser beam can be used to generate an optical lattice with
the spatially non-uniform helicoidal spin-orbit coupling. By adjusting the intensity and phase of the laser,
the spin coupling strength $\alpha$ and helical rate $\kappa$ in the system can be controlled. This setup allows
the study of nonlinear localized wave excitations in the helicoidal SOC-BEC, including stripe solitons, beating solitons, and BSSs.

In this work, we have not considered the effects of the Zeeman splitting (this effect breaks the integrability of system (\ref{helicoidal SOC})). As continuation of the work, we will conduct approximate analytical studies for BSSs in system (\ref{helicoidal SOC}) that includes the Zeeman splitting, using methods such as the multiple-scale expansion and variational approximation. Additionally, external potentials (e.g., harmonic-oscillator, parity-time-symmetric, and optical-lattice ones) significantly impact the stability of solitons and the excitation of novel soliton modes. We also plan to further investigate BSSs under the influence of external potentials.


\begin{acknowledgments}
This work was supported by the Natural Science Foundation of Hubei Province of China (Grant No. 2023AFB222) and the National Natural Science Foundation of China (Grant Nos. 11975172, 12261131495 and 12381240286).
\end{acknowledgments}

\appendix

\section{Vector beating stripe solitons (BSSs)}


Equation~(\ref{helicoidal SOC}) is gauge-equivalent to the Manakov system,
\begin{equation}
i\mathbf{u}_{t}+\frac{1}{2}\mathbf{u}_{xx}+s(\mathbf{u}^{\dag }\mathbf{u})%
\mathbf{u}=0,~~\mathbf{u}=(u_{1},u_{2})^{T},  \label{Manakov}
\end{equation}%
through the following transformation~\cite{Kartashov2017,Ding2024}:
\begin{equation}
\begin{aligned} \mathbf{\Psi }=\mathbf{Tu}=\begin{pmatrix} \nu
_{+}e^{-i(k_{\text{m}}+\kappa )x} & \nu _{-}e^{i(k_{\text{m}}-\kappa )x}\\
\nu _{-}e^{-i(k_{\text{m}}-\kappa )x} & -\nu _{+}e^{i(k_{\text{m}}+\kappa
)x} \end{pmatrix}\mathbf{u}, \end{aligned}  \label{trans1}
\end{equation}%
where $k_{\text{m}}=\sqrt{\alpha ^{2}+\kappa ^{2}}$ is the effective
momentum of the lowest-energy states, and $\mathbf{T}$ is the spatially non-uniform transformation matrix.

Here, we emphasize that, in the context of two-component BECs, systems comprising two distinct atomic species
or different hyperfine states of the same atom can be described by the coupled Gross-Pitaevskii equations
(cGPEs) \cite{Kevrekidis2015aa}, which account for varying intra- and inter-species interaction strengths.
When these strengths are equal, and no external potential is present, these cGPEs amount to the Manakov system (\ref{Manakov}).

The gauge-equivalent relationship allows one to construct solutions to Eq.~(%
\ref{helicoidal SOC}) on the basis of the exact solutions of the Manakov
system.

The Lax pair of the Manakov system is taken as~\cite{Guo2012}
\begin{equation}
\begin{aligned}
\boldsymbol{\Phi_x}=&\boldsymbol{M\Phi},~~\boldsymbol{M}\equiv i(\lambda
\boldsymbol{\sigma}+\boldsymbol{U}),\\
\boldsymbol{\Phi_t}=&\boldsymbol{N\Phi},~~\boldsymbol{N}\equiv
i(\lambda^2\boldsymbol{\sigma}+\lambda
\boldsymbol{U})+\frac{1}{2}\boldsymbol{\sigma}(\boldsymbol{U_x}-i%
\boldsymbol{U}^2), \end{aligned}  \label{A1}
\end{equation}%
where
\begin{equation}
\begin{aligned} \boldsymbol{U}=\begin{pmatrix} 0 & su_1^* & su_2^* \\ u_1 &
0 & 0 \\ u_2 & 0 & 0
\end{pmatrix},~~\boldsymbol{\sigma}=\text{diag}(1,-1,-1). \end{aligned}
\end{equation}

To derive the BSSs, we begin with the plane waves, taken as the zero-order
seed solutions:
\begin{equation}
\begin{aligned} u_1^{(0)}=&ae^{i\theta_1},~~u_2^{(0)}=0, \end{aligned}
\label{plane wave-zero}
\end{equation}%
where
\begin{equation}
\begin{aligned} \theta_j=k_jx+(sa^2-\frac{1}{2}k_j^2)t. \end{aligned}
\end{equation}%
The substitution of the above-mentioned seed solutions~(\ref{plane wave-zero}%
) in the Lax pair~(\ref{A1}) gives rise to the fundamental eigenfunction
solution,
\begin{equation}
\begin{aligned}
\boldsymbol{\Phi_1}=(\phi,\varphi,\psi)^T=GHKLe^{-i[\lambda_1
x+(sa^2+\lambda_1^2)t]}, \end{aligned}  \label{eigenfunction}
\end{equation}%
where $G=\text{diag}(1,ae^{i\theta _{1}},e^{i\theta _{2}})$, $%
L=(l_{1},l_{2},l_{3})^{T}$, $K=\text{diag}(e^{-iA(\mu _{1})},e^{-iA(\mu
_{2})},e^{-iA(k_{2})})$, and
\begin{equation}
\begin{aligned} H=\begin{pmatrix} 1 & 1 & 0 \\ \frac{1}{k_1-\mu_1} &
\frac{1}{k_1-\mu_2} & 0 \\ 0 & 0 & 1 \end{pmatrix}, \end{aligned}
\end{equation}%
with $l_{j}$~$(j=1,2,3)$ representing the characteristic
parameters of the initial soliton and $A(\mu )=\mu (x-%
\frac{\mu }{2}t)$.

Then, the fundamental BSS solutions of Eq.~(\ref{helicoidal SOC}) can be
derived by utilizing the gauge transformation~(\ref{trans1}) and the
following Darboux transformation of Manakov system
\begin{equation}
\begin{aligned}
\boldsymbol{U}[1]=\boldsymbol{U}+\frac{\lambda_1^*-\lambda_1}{%
\boldsymbol{y_1}^{\dag} \boldsymbol{\Lambda y_1}}
[\boldsymbol{y_1}\boldsymbol{y_1}^{\dag}\boldsymbol{\Lambda},\boldsymbol{%
\sigma}], \end{aligned}  \label{DT1}
\end{equation}%
where $\boldsymbol{y_{1}}\equiv v_{1}(x,t)\boldsymbol{\Phi _{1}}$ with $%
v_{1}(x,t)$ being a nonzero complex function, $\boldsymbol{\Lambda }=\text{%
diag}(1,s,s)$ and commutator of quantum operators is fixed as $[A,B]=AB-BA$.
Substituting the eigenfunction~(\ref{eigenfunction}) into the Darboux
transform~(\ref{DT1}) with $v_{1}(x,t)=e^{i[\lambda _{1}x+(sa^{2}+\lambda
_{1}^{2})t]}$ and combining the gauge transformation~(\ref{trans1}), we can
obtain the BSS solutions of Eq.~(\ref{helicoidal SOC}) as explicitly shown
by Eq.~(\ref{dark-bright soliton}).

\section{$N$-th-order beating stripe solitons (BSSs)}

Utilizing the plane-wave zero-order seed solutions~(\ref{plane wave-zero})
as the starting point, we can derive the $N$-th-order BSS solutions of Eq.~(%
\ref{helicoidal SOC}) by means of the gauge transform~(\ref{trans1}) and
application of the following $N$-th-order Darboux transform, which is
specific to the Manakov system:
\begin{equation}
\begin{aligned}
\boldsymbol{U}[N]=\boldsymbol{U}+[\boldsymbol{\sigma},\boldsymbol{Y}%
\boldsymbol{\Sigma}^{-1} \boldsymbol{Y}^{\dag}\boldsymbol{\Lambda}],
\end{aligned}  \label{DTN}
\end{equation}%
where $\boldsymbol{Y}=(\boldsymbol{y_{1}},\boldsymbol{y_{2}},\ldots ,%
\boldsymbol{y_{N}})$ with $\boldsymbol{y_{j}}\equiv v_{j}(x,t)\boldsymbol{%
\Phi _{j}}$, $\boldsymbol{\Phi _{j}}$ is the eigenfunction solution of Lax
pair~(\ref{A1}) at $\lambda =\lambda _{j}$ with the plane-wave zero-seed
solutions~(\ref{plane wave-zero}), and $\boldsymbol{\Sigma }=(\Sigma
_{ij})_{N\times N}$ with $\Sigma _{ij}=\boldsymbol{y_{i}}^{\dag }\boldsymbol{%
\Lambda }\boldsymbol{y_{j}}/(\lambda _{j}-\lambda _{i}^{\ast })$. The
eigenfunction solution $\boldsymbol{\Phi _{j}}$ has the form
\begin{equation}
\begin{aligned} \boldsymbol{\Phi_j}=GH_jK_jL_je^{-i[\lambda_j
x+(sa^2+\lambda_j^2)t]}, \end{aligned}  \label{eigenfunctionj}
\end{equation}%
where $G=\text{diag}(1,ae^{i\theta _{1}},e^{i\theta _{2}})$, $%
L_{j}=(l_{j1},l_{j2},l_{j3})^{T}$, $K_{j}=\text{diag}(e^{-iA(\mu
_{j1})},e^{-iA(\mu _{j2})},e^{-iA(k_{2})})$, and
\begin{equation}
\begin{aligned} H_j=\begin{pmatrix} 1 & 1 & 0 \\ \frac{1}{k_1-\mu_{j1}} &
\frac{1}{k_1-\mu_{j2}} & 0 \\ 0 & 0 & 1 \end{pmatrix}, \end{aligned}
\end{equation}%
with $l_{jk}$~$(k=1,2,3)$ being complex parameters, and $A(\mu )=\mu (x-%
\frac{\mu }{2}t)$. The eigenvalues $\mu _{j1}$ and $\mu _{j2}$ are solutions
of the following quadratic equation with spectral parameter $\lambda _{j}$
\begin{equation}
\begin{aligned} \mu^2+(2\lambda_j-k_1)\mu-sa^2-2k_1\lambda_j=0. \end{aligned}
\label{quadratic equationj}
\end{equation}%
Substituting the $N$-th-order solutions~(\ref{DTN}) into gauge transform~(%
\ref{trans1}), we can obtain the $N$-th-order BSSs for the original Eq.~(\ref%
{helicoidal SOC}).


\end{document}